\documentclass[10pt, twocolumn, conference]{IEEEtran}
\IEEEoverridecommandlockouts

\usepackage{subfigure}
\usepackage{paralist}
\usepackage{color}
\usepackage{url}
\usepackage[linesnumbered,ruled,boxed,vlined]{algorithm2e}

\usepackage{amsmath}
\usepackage{amssymb}
\usepackage{graphicx}
\usepackage{bm}
\usepackage{paralist}
\usepackage{subfigure}
\usepackage{times}
\usepackage{balance}
\usepackage{cite}
\usepackage{comment}

\newcommand{\separator}{
  \begin{center}
    \rule{\columnwidth}{0.3mm}
  \end{center}
}

\def\Bl{\Bigl}
\def\Br{\Bigr}

\def\eg{{\it e.g.}}
\def\ie{{\it i.e.}}

\newtheorem{theorem}{Theorem}

\newtheorem{proposition}{Proposition}
\newtheorem{lemma}{Lemma}
\newtheorem{definition}{Definition}

\newcommand{\bprob}[1]{\mathbb{P}\Bl[ #1 \Br]}
\newcommand{\prob}[1]{\mathbb{P}[ #1 ]}

\newcommand{\beq}{\begin{eqnarray*}}
\newcommand{\eeq}{\end{eqnarray*}}
\newcommand{\beqn}{\begin{eqnarray}}
\newcommand{\eeqn}{\end{eqnarray}}
\newcommand{\bemn}{\begin{multiline}}
\newcommand{\eemn}{\end{multiline}}

\newcommand{\note}[1]{{\color{red}{[\textit{#1}}]}}

\newcommand\etal{{\em et al.}}

\allowdisplaybreaks
\begin{document}

\title{Information Source Finding in Networks:\\ Querying with Budgets 
              }

\author
{Jaeyoung Choi$^*$, Sangwoo Moon$^\dag$, Jiin Woo$^{\dag\dag}$, Kyunghwan Son$^\dag$, Jinwoo Shin$^\dag$ and Yung Yi$^\dag$
\thanks{ $^*$: Department of AI Software, Gachon University,
Republic of Korea. (e-mails: jychoi19@gachon.ac.kr).  $\dag$: Department of Electrical Engineering, KAIST,
Republic of Korea. (e-mails:\{mununum, kevinson9473, jinwoos, yiyung\}@kaist.ac.kr). $\dag\dag$: NAVER cooperation, Republic of Korea. (e-mails: jidoll4366@gmail.com). Yung Yi is the corresponding author. Part of this work was presented at the IEEE INFOCOM 2017 \cite{Choi17} and IEEE ISIT 2018 \cite{Choi18}.}
}

\maketitle

\begin{abstract}
In this paper, we study a problem of detecting the source of diffused information by querying individuals, given a sample snapshot of the information diffusion graph, where two queries are asked:
{\em (i)} whether the respondent is the source or not, and {\em (ii)} if not, which neighbor spreads the information to the respondent.
We consider the case when respondents may not always be truthful and some cost is taken for each query.
Our goal is to quantify the necessary and sufficient budgets to achieve the detection probability $1-\delta$ for any given $0<\delta<1.$ To this end, we study
two types of algorithms: adaptive and non-adaptive ones, each of which corresponds to whether
we adaptively select the next respondents based on the
answers of the previous respondents or not.
We first provide the information theoretic lower bounds for the necessary budgets in both algorithm types.
In terms of the sufficient budgets, we propose two practical estimation algorithms, each of non-adaptive and adaptive types, and for each algorithm,
we quantitatively analyze the budget which ensures $1-\delta$ detection accuracy. This theoretical analysis not only quantifies the budgets needed by practical estimation algorithms achieving a given target detection accuracy in finding the diffusion source, but also
enables us to quantitatively characterize the amount of extra budget required in non-adaptive type of estimation, refereed to as {\em adaptivity gap}.
We validate our theoretical findings over synthetic and real-world social network topologies.
\end{abstract}




\section{Introduction}

Information spread in networks is universal to model many real-world
phenomena such as propagation of infectious diseases, diffusion of
a new technology, computer virus/spam infection in the Internet,
and tweeting and retweeting of popular topics. The problem of finding the information source is to
  identify the true source of information spread. This is clearly
  of practical importance, because harmful diffusion
  can be mitigated or even blocked, \eg, by vaccinating humans or installing
  security updates \cite{Kai2016}.
  Recently, extensive research attentions for this problem
have been paid for various network topologies and diffusion models
\cite{shah2010,shah2012,zhu2013,Luo2013,bubeck2014,Kai2016, Chang2015,farajtabar2015}, whose major
interests lie in constructing an efficient estimator and providing
theoretical analysis on its detection performance.

Prior work has shown that the
  detection probability cannot be beyond 31\% even for regular trees if
  the number of infected nodes is sufficiently large.
This directly or indirectly conclude that the information source finding turns out to be a challenging task unless sufficient side information
or multiple diffusion snapshots are provided.
There have been several research efforts which use multiple snapshots \cite{Zhang2014} or a side information about a restricted superset the true source belongs \cite{dong2013}, thereby the detection performance is significantly improved. Another type of side information is the one obtained from {\em querying}, \ie, asking questions to a subset of infected nodes and gathering more hints about who would be the true information source \cite{Choi17}. The focus of this paper is also on querying-based approach (we will shortly present the difference of this paper from \cite{Choi17} at the end of this section).

In this paper, we consider an \emph{identity with direction} (id/dir in short) question as follows. First, a querier asks an identity question of
whether the respondent \footnote{In this paper, we call `respondent' by the node who is asked a question from the querier.} is the source or not, and if ``no", the respondent is subsequently asked the direction question of  which neighbor
spreads the information to the respondent. Respondents may be untruthful with some
probability so that the multiple questions to the same respondent are allowed to filter the untruthful answers, and the total number of questions can be asked within a given budget.
We consider two types of querying schemes: {\em (a) Non-Adaptive ({\bf NA})} and {\em (b) ADaptive ({\bf AD}).} In {\bf NA}-querying,
    a candidate respondent set is first chosen,
    and the id/dir queries are asked in a batch manner.
    In {\bf AD}-querying, we start with some
initial respondent, iteratively ask a series of id/dir questions to the current respondent, and adaptively determine the next respondent using the
(possibly untruthful) answers from the previous respondent, where this iterative
querying process lasts until the entire budget is used up.
In general, it is known that the adapitve manner of processing data from observation or sampling is more efficient than that of the non-adaptive one in other research domains such as seeding problem in social network \cite{Fujii19, Singer16}, crowdsourcing problem \cite{Sewoong16} and community detection \cite{Yun14}. 
In our work, we first quantify the adaptiveness gain using the queried data on information source detection problem.

We summarize our main contributions in what follows.

\smallskip
\begin{compactenum}[(a)]
\item First, we formulate an
  optimization problem that maximizes the detection probability over the
  number of questions to be asked, the
  candidate respondent set, and the estimators for a given diffusion
  snapshot and the answer samples. We discuss analytical challenges of the problem and then propose a tractable and appropriate model that characterizes upper and lower bounds of the detection probability. Under this model, we obtain the necessary budgets for both querying schemes to achieve the $(1-\delta)$ detection probability for any given $0<\delta<1.$
To this end, we establish information theoretical lower bounds
from the given diffusion snapshot and the answer samples from querying.
Our results show that it is necessary to use the budget $\Omega \left(\frac{(1/\delta)^{1/2}}{\log(\log(1/\delta))}\right)$
for the {\bf NA}-querying, whereas $\Omega \left(\frac{\log^{1/2} (1/\delta)}{\log(\log(1/\delta))}\right)$
for the {\bf AD}-querying, respectively.

\item Second, to obtain the sufficient amount of budget for $(1-\delta)$ detection performance, we consider two estimation algorithms, each for both querying schemes, based on a simple Majority Voting (MV) to handle the untruthful answer samples. We analyze simple, yet powerful estimation algorithms whose time complexities are 
$O(\max\{N,K^2\})$ for {\bf NA}-querying and $O(\max\{N,K\})$ for {\bf AD}-querying, 
respectively, 
where budget is $K$ and the number of infected nodes is $N.$ 
Our results show that the sufficient query budgets are  
$O \left(\frac{(1/\delta)}{\log(\log(1/\delta))}\right)$
for the {\bf NA}-querying and 
$O \left(\frac{\log^2 (1/\delta)}{\log(\log(1/\delta))}\right)$
for the {\bf AD}-querying. 
The gap between necessary and sufficient budgets in both querying schemes is due to our consideration of simple, yet practical estimation algorithms based on majority voting, caused by the fact that the classical ML (Maximum Likelihood)-based estimation is computationally prohibitive and even its analytical challenge is significant.
Our quantification of necessary and sufficient budgets enables us to
obtain the lower and upper bounds of the \emph{adaptive gap}, \ie, the gain of adaptive querying scheme compared to non-adaptive one. 

\item Finally, our analytical results above provide useful guidelines on
  how much budget is necessary and sufficient  to guarantee a given detection performance
  for different querying types when users are untruthful. We validate
  our findings via extensive simulations over popular random graphs
  (\emph{Erd\"{o}s-R\'{e}nyi}  and scale-free graphs) and a real-world
  Facebook network. As an example, in Facebook network,
the {\bf AD}-querying requires about 600 queries to achieve above 
90\% detection probability when $p=0.6$ and $q=0.3$ (Here, $p$ and $q$ are the probabilities of telling the truth for id question and direction question, respectively.) whereas the {\bf NA}-querying requires more than 4000 queries to achieve same detection probability. This indicates that adaptive use of the querying budget is more efficient for finding the source than that of non adaptive one as we expected in the analytical results. 
\end{compactenum}

\smallskip
The remainder of this paper is organized as follows. Section \ref{sec:related} discusses related literature. In section 
\ref{sec:model}, we introduce our querying model under the information diffusion and describe our goal of the paper. The theoretical results for {\bf NA}-querying and {\bf AD}-querying with their adaptive gap will be presented in section \ref{sec:noninteractive} and \ref{sec:interactive}, respectively and the corresponding proof will be provided in section \ref{sec:proof}. In section \ref{sec:numerical}, we depict the simulation results and conclude the paper in section \ref{sec:conclusion}. 

\section{Related Work}
 \label{sec:related} 
The research on rumor\footnote{We use the terms ``information'' and ``rumor'' interchangeably.} source detection has recently received
significant attentions. In this section, we divide them into the following two categories: 
(1) estimation of a single source and (2) estimation of multiple sources.


\smallskip
\noindent{\bf \em (1) Single source estimation.} 
Shah and Zaman
\cite{shah2010,shah2012,shah2010tit}  first considered the single 
source detection problem over a connected network. 
They introduced the metric
called {\em rumor centrality} --- a simple topology-dependent
metric for a given diffusion snapshot. They showed that the rumor centrality describes the likelihood
function when the underlying network is a regular tree and the diffusion follows
the Susceptible-Infected (SI) model. Zhu and Ying
\cite{zhu2013} solved the source detection problem under the Susceptible-Infected-Removed (SIR) model and took a sample path approach to
solve the problem, where the {\em Jordan center} was used,
being extended to the case of sparse observations \cite{zhu2014}. All the above detection mechanisms
correspond to point estimators, whose detection performance tends to be
low.  

There were several attempts to boost up the detection probability.
Wang \etal \cite{Zhang2014} showed that observing multiple different
epidemic instances can significantly increase the detection
probability. Dong \etal \cite{dong2013} assumed that there exist 
a restricted set of source candidates, where they showed the increased
detection probability based on the Maximum a Posterior Estimator (MAPE).
Choi \etal \cite{Choi2016} showed that the anti-rumor spreading under
some distance distribution of rumor and anti-rumor sources helps to find
the rumor source by using the MAPE.
Recently, there have been some
approaches for a realistic graph topology or a partial observation setting. Luo \etal \cite{Leng2014} considered the problem of estimating an infection
source for the SI model, in which not all infected
nodes can be observed. When the network is a tree, they showed
that an estimator for the source node associated with the most
likely infection path that yields the limited observations is given
by a Jordan center, i.e., a node with the minimum distance to the
set of observed infected nodes. They also proposed approximate
source estimators for general networks.
Kumar \etal \cite{Kumar2017} considered the case where additional relative information about the infection times of a fraction of node pairs is also available to the estimator. The authors took a complementary approach where the estimator for general networks ranks each node based on counting the number of possible spreading patterns with a given node as root that are compatible with the observations.
Zhu \etal \cite{Kai2016} considered the \emph{Erd\"{o}s-R\'{e}nyi} (ER) random graph and proposed a
new source localization algorithm, called the Short-Fat Tree
(SFT) algorithm. The algorithm selects the
node such that the Breadth-First Search (BFS) tree from the
node has the minimum depth but the maximum number
of leaf nodes. Performance guarantees of SFT under the
Independent Cascade (IC) model have been established for both tree
networks and the ER random graph. Chang \etal \cite{Chang2015} considered both infected and uninfected nodes to estimate the likelihood for the detection on a general loopy graph. They considered a MAPE to detect the source for the general graph using the rumor centrality as a prior.  Jiang \etal \cite{Jiao2018} considered time-varying networks where there is the temporal variation in the topology of the underlying networks. They introduced an effective approach used in criminology to overcome the challenges. 


\smallskip
\noindent{\bf \em (2) Multiple sources estimation.} 
In the multiple sources estimation problem, it is required to infer the set of source nodes which results in the diffusion snaps
hot. Despite the difficulty of the problem, some prior studies tried to solve this problem by appropriate set estimation methods. 
Prakash \etal \cite{ICDM17} proposed to employ the Minimum Description Length (MDL) principle to identify the best set of seed nodes and virus propagation ripple, which describes the infected graph most succinctly.
They proposed a highly efficient algorithm to identify likely sets of seed nodes
given a snapshot and show that it can optimize the
virus propagation ripple in a principled way by maximizing the likelihood. Zhu \etal \cite{ZhuAAAI17} proposed a new source localization algorithm, named Optimal-Jordan-Cover (OJC). The
algorithm first extracts a subgraph using a candidate selection algorithm that selects source candidates
Considering the heterogeneous
SIR diffusion in the ER random graph, they proved
that OJC can locate all sources with probability one asymptotically with partial observations. Ji \etal \cite{JI17} developed a theoretical framework to estimate rumor sources, given an observation of
the infection graph and the number of rumor sources. They considered a scenario that sources start to spread the rumor at a
different time. They first studied the two-source identification problem
in a tree network under the SI model and then generalized the framework to multiple sources and for
all graph types. Jaing \etal \cite{Jaing15} proposed a novel method to identify multiple
diffusion sources, which can address the question of how many sources there are and where the diffusion
emerges. They derived an optimization formulation for the multi-source identification
problem and quantified the detection performance for the proposed algorithm.

To the best of our knowledge, our paper is the first to quantitatively consider
the querying approach, which uses the snapshot of the infection graph and additional side-information from querying, appropriately. The results give the necessary and sufficient amount of querying budgets to achieve the target detection probability based on information theoretical techniques and simple algorithms with low complexity (Majority Voting based one), respectively.

\section{Model and Goal}
\label{sec:model}
\subsection{Diffusion Model and MLE}

\noindent{\bf \em Diffusion Model.}
We consider an undirected graph
$G=(V,E),$ where $V$ is a countably infinite set of nodes and $E$ is
the set of edges of the form $(i,j)$ for $i, j\in V$.  Each node
represents an individual in human social networks or a computer host
in the Internet, and each edge corresponds to a social relationship
between two individuals or a physical connection between two Internet
hosts. As an information spreading model, we consider a SI model
under exponential distribution with rate of $\lambda_{ij}$ for the edge $(i,j),$ and all nodes are initialized to be
susceptible except the information source. Once a node $i$ has an information,
it is able to spread the information to another node $j$ if and only if
there is an edge between them. We
denote by $v_1 \in V$ the information source, which acts as a node that
initiates diffusion and denote by $V_N \subset V$, $N$ infected nodes
under the observed snapshot $G_N\subset G$.
In this paper, we consider the case when $G$ is a regular tree, the diffusion rate $\lambda_{ij}$ is homogeneous with unit rate, \ie, $\lambda_{ij}=\lambda =1 $, and $N$ is
large, as done in many prior work
\cite{shah2010,shah2012,Khim14,dong2013,Zhang2014}.
We assume that there is no prior distribution about the source, \ie,
the uniform distribution.

\smallskip
\noindent{\bf \em Maximum Likelihood Estimator (MLE).}
As a preliminary, we explain the notion of {\em rumor centrality,} which is a
graph-theoretic score metric and is originally used in detecting the
rumor source in absence of querying and users' untruthfulness. This notion is also importantly used in our framework
as a sub-component of the algorithms for both {\bf NA}-querying and
{\bf AD}-querying.
In regular tree graphs, Shah and Zaman \cite{shah2010} showed that the
source chosen by the MLE becomes the node
with highest rumor centrality. Formally, the estimator chooses $v_{RC}$ as
the rumor source defined as
\begin{eqnarray}\label{eqn:ML}
       \vspace{-0.4cm}
v_{RC} &=& \arg\max_{v \in V_N} \mathbb{P}(G_N | v=v_1)\cr
       &=&\arg\max_{v \in V_N} R(v, G_N),
              \vspace{-0.4cm}
\end{eqnarray}
where $v_{RC}$ is called \emph{Rumor Center} (RC) and $R(v, G_N)$ is the
rumor centrality of a node $v$ in $V_N$. The rumor centrality of a
particular node is calculated only by understanding the graphical
structure of the rumor spreading snapshot, \ie, $R(v, G_N) =N!
\prod_{u \in V_N} (1/T^v_u)$ where $T^v_u$ denotes the number of nodes
in the subtree rooted at node $u$, assuming $v$ is the root of tree
$G_N$ (see \cite{shah2010} for details).


\subsection{Querying Model}


\noindent{\bf \em Querying with untruthful answers.}
Using the diffusion snapshot of the information, a detector performs querying which refers to a process of asking some questions. We assume that a fixed budget $K$
is given to the detector (or the querier) and a unit budget has worth of asking one pair of id/dir question, \ie,
 ``Are you the source?'' first and if the respondent answers ``yes" then it is done. Otherwise, the detector subsequently asks a direction question as ``Which neighbor spreads the information to you?''.
  In answering a query, we consider that each respondent $v$
  is only partially truthful in answering id and dir questions, with probabilities of being truthful, $p_v$ and $q_v$, respectively. 
To handle untruthful answers, the querier may ask to a respondent $v$ the question multiple times, in which $v$'s truthfulness is assumed to be independent. More precisely, we define a Bernoulli random variable $I$ that represents a respondent's answer for the id question, such that $I$ is one with probability (w.p.) $p_v$ and $I$ is zero w.p. $1-p_v$.
  Similarly, we define a respondent's random answer $D$ for the dir question, such that $D$ is one w.p. $q_v$ and $D$ is $i$ w.p. $(1-q_v)/(d-1)$, for $i = 2, \ldots, d.$
  We also assume that homogeneous truthfulness across individuals, i.e., $p_v = p$ and $q_v = q$ for all $v \in V_N$ and $p > 1/2, q>1/d$ meaning that all answers are more biased to the truth.  
 To abuse the notation, we use $H(p)$ and $H(q)$ to refer to the entropies of $I$ and $D$, respectively. Throughout this paper, we also use the standard notation $H(\cdot)$ to denote the entropy of a given random variable or vector.

  In terms of querying schemes, we consider the following two types of querying: {\em non-adaptive} (NA) and {\em adaptive} (AD) querying.

\begin{figure}
\begin{center} \centering
\subfigure[Non-adaptive (NA)-querying.]{\includegraphics[width=0.23\textwidth]{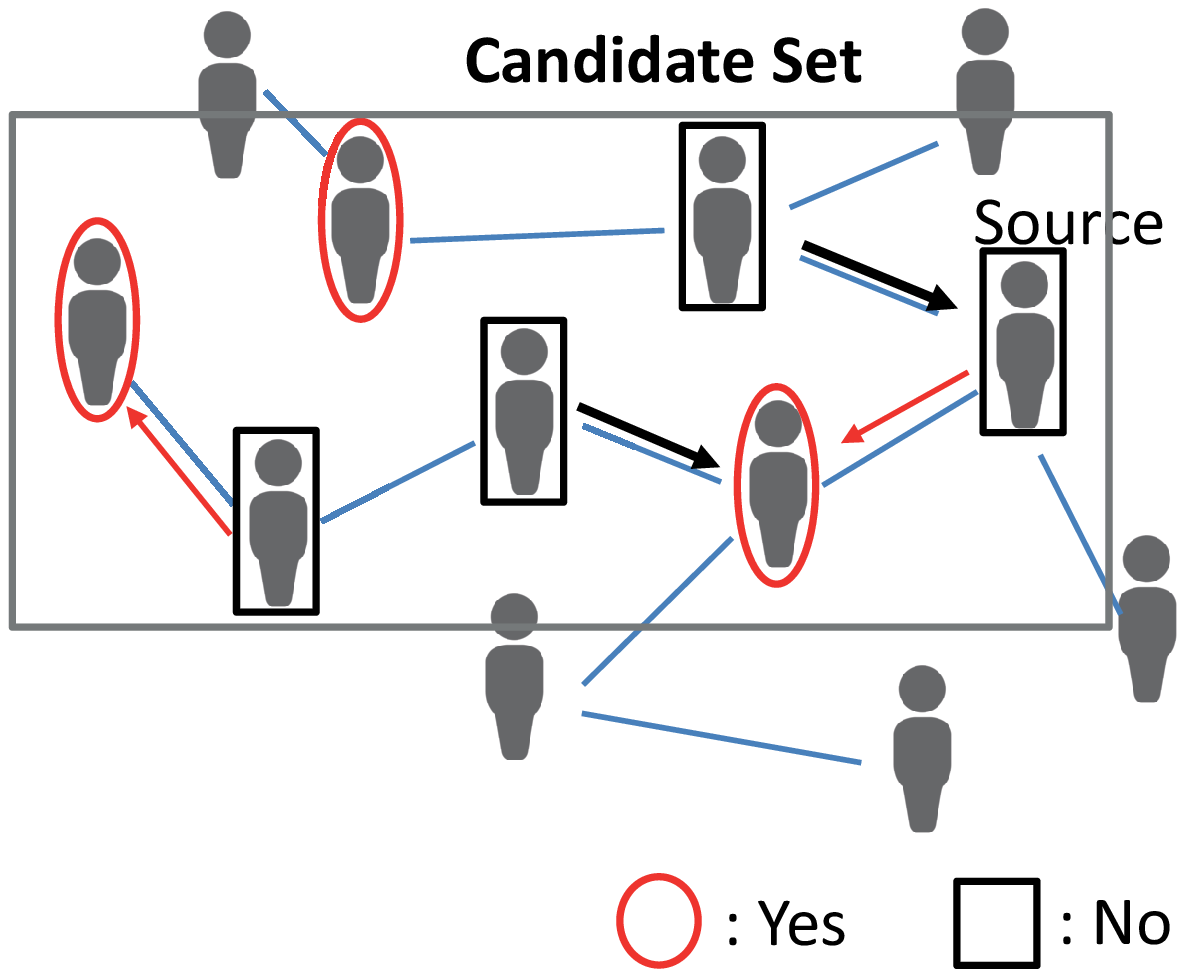}\label{fig:Noninter}}
\subfigure[Adaptive (AD)-querying.]{\includegraphics[width=0.23\textwidth]{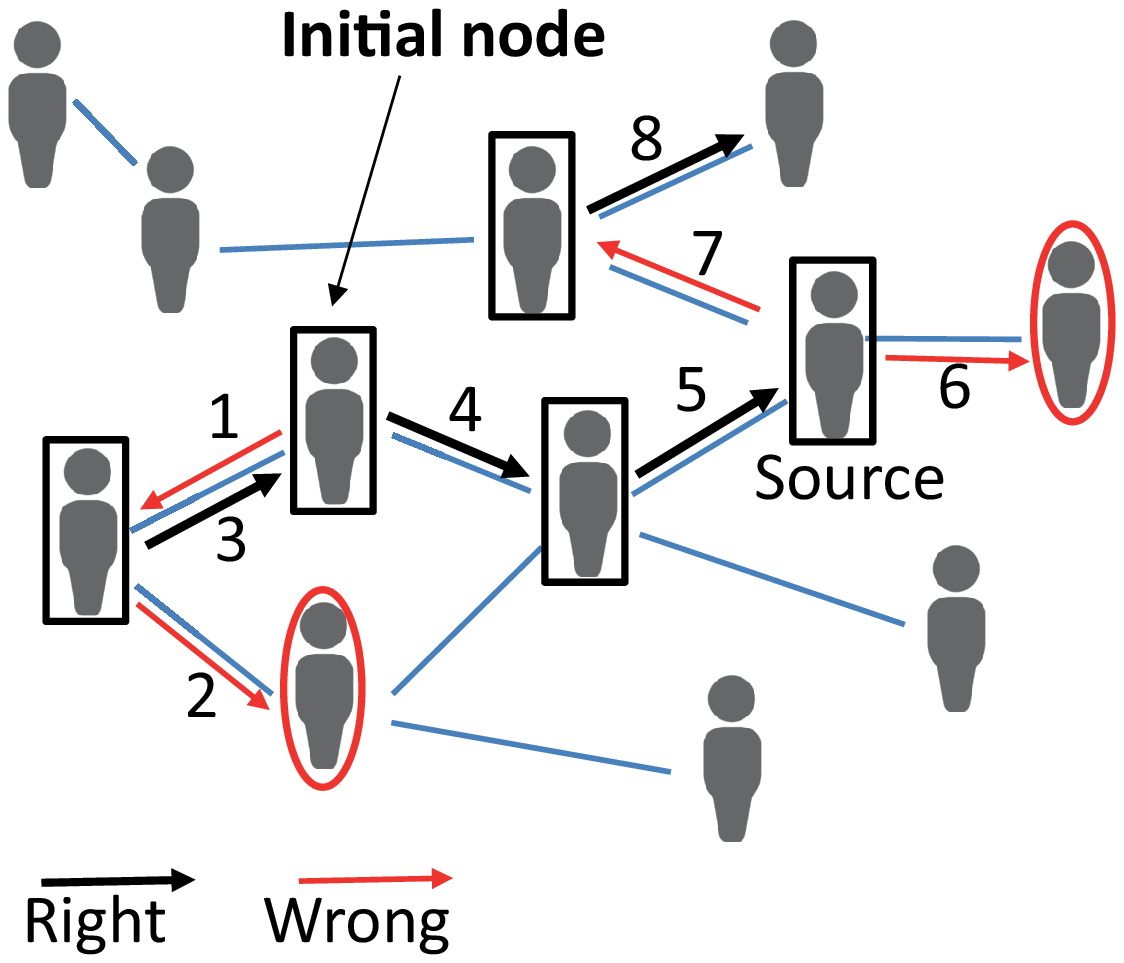}\label{fig:Inter}}
\hspace{0cm}
\end{center}
\caption{Examples of two querying types with untruthful
  answers ($r=1$). In (a), the querier selects a candidate set (a large square) and asks just one id/direction question in a batch manner under the untruthful answers. In (b),
  starting from the initial node, the querier first asks one id/direction question and adaptively tracks the true source with the untruthful answers. (In (b), ``True" is the direction of true parent and ``Wrong" is the wrong direction.)}
\label{fig:querying}
 \vspace{-0.2cm}
\end{figure}




\smallskip
\noindent {\bf \em NA-querying.} 
In this querying, it is parameterized by
  $r$, where a querier first chooses $K/r$  ($K$ is a multiple of $r$ for expositional convenience)
  candidate nodes in a batch which is believed to contain the true source, then ask id/dir question to each respondent inside the candidate set $r$ times. We call $r$ the {\em repetition count} throughout this paper. Then, the querier finally run an estimation algorithm based on the answers from all the respondents.  
  



\smallskip
\noindent {\bf \em AD-querying.}
In the adaptive querying, a querier first chooses an initial node to ask the id/dir question, possibly multiple times, and
the querier adaptively determines the next respondent using the answers from the previous respondent, which is repeated until the entire budget is exhausted. The number of selected nodes in this query is also $K/r$ and some nodes can be selected multiple times by the querier. Then, the querier finally run an estimation algorithm based on the answers from all the respondents.

\smallskip
In {\bf NA}-querying, Fig.~\ref{fig:Noninter} illustrates a
  candidate set of nodes inside a square, id/dir querying is performed in a batch with $r=1$. In {\bf AD}-querying, Fig.~\ref{fig:Inter} shows an example scenario that
  starting from the initial node, a sequence of nodes answer the
  queries truthfully or untruthfully for $r=1$.

\subsection{Goal}
 As a performance metric,
 we consider the required number of budget $K$ to achieve the detection probability at least $1-\delta$
 for a given $0<\delta <1$.
Then, our goal is to answer for the following questions.

\smallskip
\begin{compactenum}[Q1.]
\item  \emph{How many queries are necessary and sufficient for {\bf NA}-and {\bf AD}-querying?} 
We first consider an MLE of the snapshot and querying data as a target estimator for both querying scenarios, and discuss its technical complexities. 
This motivates us to consider a restricted, but still wide class of algorithms, where we establish information theoretical lower bound to obtain the necessary number of budgets and study a sufficient number of budgets by analyzing the performance of an MV-based estimation. 



\smallskip
\item  \emph{How large is the adaptivity gain?}
We finally quantify how much adaptivity on the querying scheme
 enriches the detection probability of the information source under the individual's untruthfulness. We define the quantification by \emph{adaptivity gap} and obtain its lower and upper bounds by using the obtained number of necessary and sufficient budgets for each querying. 
\end{compactenum}

\section{NA-Querying: Necessary and Sufficient Budgets}
\label{sec:noninteractive}

\subsection{Challenges and Algorithm Class}
\label{subsec:algorithm class NA}
In {\bf NA}-querying, we first describe the hardness of obtaining an optimal algorithm in the entire algorithm set. To do this, we first denote $C_r \subset V_N$, by the candidate set of query with size $K/r$ for a given $r.$ Then
we next describe the data of querying answer as follows.
Let $A_r (p,q):=(X(p),Y(q))$ be the answer vector where
$X(p):=[x_1 (p), x_2 (p), \ldots, x_{K/r} (p)]$ with $0 \leq x_i (p)\leq r $
representing the number of answers ``yes'' and $Y(q):=(Y_1(q),\ldots, Y_{K/r}(q))$
with $Y_{i} (q)= [y_1(q), y_2(q), \ldots, y_d(q)]$, the answer vector
for the respondent $i$, where $0\leq y_j(q) \leq r$ that represents the number
of ``designations'' to $j$-th neighbor ($1 \leq j \leq d$) of the respondent $i$.
Then, the MLE of querying as an optimal estimation algorithm is to solve the following problem:

\vspace{-0.1cm}
\separator
\vspace{-0.1cm}
\begin{align}\label{eqn:OPT}
\text{\bf OPT-NA:} &  \quad \max_{ 1 \leq r \leq K} \max_{C_{r }}
  \max_{v \in C_{r}} \bprob{G_N,A_{r}(p,q)| v=v_1},
\end{align}
where the inner-most max corresponds to the MLE that maximizes the likelihood of $G_N$ and the query answer sample $A_{r }(p,q)$ under the assumption of $v=v_1.$
\vspace{-0.3cm}
\separator
\vspace{-0.1cm}

\smallskip
\noindent{\bf \em Challenges.} We now explain the technical challenges
in solving {\bf OPT-NA} as similar in \cite{Choi17}. To that end, let us
consider the following sub-optimization in {\bf OPT-NA} for a fixed $ 1\le
r \le K$:
\begin{align}\label{eqn:subOPT}
\text{\bf SUB-OPT-NA:} &  \quad \max_{C_{r}}
  \max_{v \in C_{r}} \bprob{G_N,A_{r}(p,q) | v=v_1}.
\end{align}
Then, the following proposition provides the solution of {\bf SUB-OPT-NA}
whose proof is provided in the supplementary material. 

\smallskip
\begin{proposition}\label{pro:subOPT}
  Construct $C_r^*$ by including the $K/r$ nodes in the decreasing order
  of their rumor centralities. Then, $C_r^*$ is the solution of {\bf SUB-OPT-NA}.
\end{proposition}

\smallskip
Despite our knowledge of the solution of {\bf SUB-OPT-NA},
solving {\bf OPT-NA} requires an analytical form of the objective value of
{\bf SUB-OPT-NA} for $C_r^*$ to find the optimal repetition count, say
$r^*.$ However, analytically computing the detection probability for
a given general snapshot is highly challenging due to the following
reasons. We first note that
\begin{align}
  \label{eq:trade}
&\max_{v \in C_{r}} \bprob{G_N,A_{r}(p,q) | v=v_1} \cr
& =  \underbrace{\prob{v_1 \in C_{r}^*}}_{(a)} \times \underbrace{\max_{v \in C_r^*} \bprob{G_N,A_r (p,q)| v=v_1, v_1 \in C_{r}^*}}_{(b)}.
\end{align}
First, the term $(a)$ is difficult to analyze, because only the MLE of snapshot
 allows graphical and thus analytical characterization as
discussed in \cite{shah2010} but other
nodes with high rumor centrality is difficult to handle due to the
randomness of the diffusion snapshot.
Second, in $(b)$, we observe
that using the independence between $G_N$ and $A_{r}(p,q),$ by letting the
event $\mathcal{A}(v) = \{v=v_1, v_1 \in C_{r}^*\},$
\begin{align}
\label{eq:decomposition}
\hat{v}&= \arg \max_{v \in C_{r}^*} \bprob{G_N,A_{r}(p,q) \mid  \mathcal{A}(v)} \cr
&= \arg \max_{v \in C_{r}^*} \bprob{A_{r}(p,q)
  \mid \mathcal{A}(v)} \times \bprob{G_N\mid \mathcal{A}(v)}\cr
  &= \arg \max_{v \in C_{r}^*} \bprob{X(p)
  \mid \mathcal{A}(v)} \times \bprob{Y(q)
  \mid \mathcal{A}(v)}\cr 
  &\qquad \qquad \quad\times \bprob{G_N\mid \mathcal{A}(v)},
\end{align}
where the last equality is due to the independence of $X(p)$ and $Y(q)$.
Then, the node $\hat{v}$ maximizing $(b)$ is the node $v$ that has the
maximum {\em weighted} rumor centrality which
is hard to obtain a characterization due to the randomness of the answer for querying,
thus resulting in the challenge of computing $r$ that
maximizes the detection probability in {\bf OPT-NA}. Hence, as our consideration to handle the probability $(a)$, we restrict our focus of the following class of {\bf NA}-querying mechanisms, denoted by $\mathcal{NA}(r,K)$, in this paper:


\smallskip
\begin{definition}(Class $\mathcal{NA}(r,K)$)
\label{def:na}
In this class of {\bf NA}-querying schemes with the repetition count $r$
and a given budget $K,$
the querier first chooses the candidate set of $ K/r $ infected nodes according to the following selection rule:
We initially select the node RC and
add other infected nodes in the increasing sequence in terms of the {\em hop-distance} from the RC. Then, the querier asks the id/dir question $r$ times to each node in the selected candidate set.
\end{definition}

\smallskip
The hop-based candidate set selection has been introduced in \cite{Khim14}. In this paper, the authors obtained the probability that the information source is in a set, denoted by $H_L$, which consists of all infected nodes within the distance $L>0$ to the RC for $d$-regular tree. They showed that it is a good approximation to the optimal candidate set which maximizes the probability that the source is included. We use this result in our analysis. From the defined algorithm class $\mathcal{NA}(r,K)$, we  obtain the necessary and sufficient budgets which achieve the target detection probability as follows.

 \subsection{Necessary Budget}
\label{subsec:necessaryNA} 
We present an information theoretic lower bound
of the budget for the target detection probability $(1-\delta)$ inside the class of $\mathcal{NA}(r,K).$ We let $\mathcal{T}(r)=[T_1,T_2,\ldots, T_{ K/r }]$ be the random vector where each $T_i$ is the random variable of infection time of the $i$-th node $(1\leq i \leq K/r)$ in the candidate set. Then, by appropriately choosing $r$, we have the following theorem. 




\smallskip
\begin{theorem}
\label{theorem:lower}
Under $d$-regular tree $G$, as $N \rightarrow \infty,$ for any $0< \delta <1,$ there exists a constant $C=C(d),$ such that if
\begin{align}
\label{eqn:lower}
     K & \leq \frac{C \cdot H(\mathcal{T}(r^\star)) (2/\delta)^{1/2}}{f_{1} (d,p,q)\log(\log (2/\delta))},
    \end{align}
where
\begin{align}
f_{1} (d,p,q) & =(1-H(p))+p(1-p)(\log_2 d -H(q)), \cr
r^\star & =  \left \lfloor 1+\frac{4(1-p)\{7H(p)+2H(q)\}\log K}{3e\log(d-1)}\right\rfloor,
\end{align}
    then no algorithm in the class $\mathcal{NA}(r,K)$ can achieve the detection
probability $1-\delta.$ Here, $H(\mathcal{T}(r))$ is a function of the diffusion rate $\lambda,$ as in \cite{Sujay2012}.
\end{theorem}

\smallskip
The implications of Theorem~\ref{theorem:lower} are in order.
First, if the entropy $H(\mathcal{T}(r^\star))$ of the infection time is large,  the necessary amount of budget increases due to large uncertainty in figuring out a predecessor in the diffusion snapshot.
Especially, we have the upper bound of $H(\mathcal{T}(r^\star))$ by $H(\mathcal{T}(r^\star)) \leq \sum_{i \in C_{r^{\star}}}\ln \left(\frac{\Gamma(h_i)}{\lambda}\right)+h_i$ by using the exponential distribution of diffusion for tree network, where $h_i >0$ is the distance (hops) to the source for the node $i \in C_{r^{\star}}$. This implies that if diffusion snapshot is sparse to the source (\ie, many infected nodes have large distance to the source), the upper bound of the entropy becomes large and we need more budget.
Second, larger entropy for the answers of id/dir questions requires more
budget to achieve the target detection accuracy. Also,
when $p$ goes to $1/2$ and $q$ goes to $1/d$, \ie, no information from the querying,
results in diverging the required budget (because $f_{1}(d,p,q)$ goes to zero).
Finally, if respondents are truthful in answering for the id question (\ie, $p=1$),
the direction answers does not effect the amount of necessary budget.

\begin{algorithm}[t!]
 \caption{{\bf MVNA}$(r)$}
\label{alg:noninteractive}
{

 \KwIn{Diffusion snapshot $G_N$, budget $K$, degree $d$, truthfulness probabilities
   $p>1/2$, $q>1/d.$}
 \KwOut{Estimator $\hat{v}$ \par\noindent\hrulefill}

\smallskip

$C_r  = S_I = S_D =\emptyset$\;
Choose the candidate set $C_r$ as in Definition~\ref{def:na} and ask the id/dir questions $r$ times to each node in $C_r$\;

\For{each $v \in C_r$}{
\textbf{Step1}: Count the number of `yes'es for the
identity question, stored at $\mu(v),$ and if $\mu(v)/r \geq 1/2$ then
add $v$ to $S_I$\;
\smallskip
\textbf{Step2}: For each of $v$'s neighbors, count the number of
       designations for the dir question, choose the $v$'s neighbor, say $w$, with the largest count (under the rule of random tie breaking) as $v$'s `predecessor', and save a directed edge, called predecessor edge, $w \rightarrow v$
       \;
       \smallskip
}

Make a graph $G_{\text{pre}}$with all the predecessor edges and for each $v \in C_r,$ set $E(v) \leftarrow$ the number of all the descendants of $v$




$S_D \leftarrow \arg\max_{v\in C_r}
   |E(v)|$\;
 \If{$S_I \cap S_D=\emptyset$}{
 If $p=1$, set $ \hat{v} \leftarrow \arg\max_{v\in S_I }
   \mathbb{P}(G_N | v=v_1)$ otherwise, set
 $ \hat{v} \leftarrow \arg\max_{v\in S_I \cup S_D}
   \mathbb{P}(G_N | v=v_1)$\;
   } \Else{ $ \hat{v} \leftarrow \arg\max_{v\in S_I \cap S_D}
   \mathbb{P}(G_N | v=v_1)$\; }
 Return $\hat{v}$\;
}
\end{algorithm}

\smallskip
 \subsection{Sufficient Budget}
\label{subsec:sufficientNA} 
To compute a sufficient budget, a natural choice would be to use the MLE (Maximum Likelihood Estimator), which, however, turns out to be computationally intractable for large $N$ due to randomness of the diffusion snapshot and query answers as described by part $(b)$ in \eqref{eq:decomposition}.
Hence, to obtain the sufficient budgets, we consider a simple estimation
algorithm named {\bf MVNA}$(r)$ that is based on
majority voting for both the id and dir questions. To briefly explain how the algorithm behaves, we first select the candidate set $C_r$ of size $ K/r $ that has the least hop-distance from the RC, then we ask $r$ times of id/dir questions to each node in the candidate set (Line 2). Then, we filter out the nodes that are more likely to be the source and save them in $S_I$ (Line 4) and using the results of the dir questions, compute $E(v)$ that correspond to how many nodes in $C_r$ hints that $v$ is likely to be the source node (Lines 6 and 7).
Finally, we
choose a node with maximal likelihood in $S_I \cap S_D$ and if
$S_I \cap S_D = \emptyset,$ we simply perform the same task for $S_I \cup S_D.$
It is easy to see that the time complexity is $O(\max\{N,K^2\})$.

\smallskip
\noindent{\bf \em Rationale:} The rationale of {\bf MVNA}$(r)$ from the perspective of how we handle the
analytical challenges by an approximate manner is described as follows.
First, for the identity questions, consider the answer sample of node $v$ for $r$ questions,
  $x_v (p)~(1\leq x_v (p) \leq r),$ where one can easily check that for
  $x_v \geq r/2$ then the weight $\prob{X(p)| v=v_1}$ becomes
  larger than that for $x_v (p) < r/2$ due to $p>1/2.$ We use an
  approximated version with the weight from the answer samples by setting
  $\prob{X(p)|v=v_1}=1$ if $x_v (p) \geq r/2,$ and
  $\prob{X(p)|v=v_1}=0$ if $x_v (p) < r/2$.
  For the direction questions,
  we see that $\prob{Y(q)|v=v_1}=1$ for the maximum consistent edge node
  and $\prob{Y(q)|v=v_1}=0$, otherwise. Hence, this is
  two step $\{0,1\}$-weighted algorithm instead of using MLE of answer data.

Now, Theorem~\ref{theorem:noninteractive} quantifies the amount of querying
budget that is sufficient to obtain arbitrary detection probability by
appropriately choosing the number of questions to be asked.

\smallskip
\begin{theorem}
\label{theorem:noninteractive}
For any $0< \delta <1,$ the detection
probability under $d$-regular
tree $G$ is at least $1-\delta,$ as $N \rightarrow \infty,$ if
\begin{align}
\label{eqn:elower}
     K & \geq \frac{12d/(d-2)(2/\delta)}{f_{2} (d,p,q)\log(\log (2/\delta))},
    \end{align}
    where $f_{2} (d,p,q)=3(p-1/2)^2 + \frac{(d-1)p(1-p)}{3d} (q-1/d)^2 $ under {\bf MVNA}$(r^\star),$ where
    \begin{align*}
r^\star &= \left \lfloor 1+\frac{2(1-p)\{1+(1-q)^2\}\log K}{e\log(d-1)}\right\rfloor.
\end{align*}
\end{theorem}

\smallskip

We briefly discuss the implications of the above theorem.
First, we see that $(1/\delta)^{1/2}$ times more budget is required that the necessary one, which is because we consider a simple, approximate estimation algorithm.
Second, the dir question does not effect the
sufficient budget $K$ if $p=1$ \ie, no untruthfulness
for the id question as in Theorem~\ref{theorem:lower}. However, if $p<1$,
the information from the answers for the dir questions reduces the sufficient
amount of budget, because $f_{2}(d,p,q)$ increases in the denominator of \eqref{eqn:elower}. Third, we see that if $q=1/d$ the result matches that of batch query in \cite{Choi17}.\footnote{This is because the answer does not give any information of direction.} 
Finally, when $p$ goes to $1/2$ and $q$ goes to $1/d$, the required budget
diverges due to the lack of information from the querying.




\section{AD-Querying: Necessary and Sufficient Budgets}
\label{sec:interactive}

In the {\bf AD}-querying, unlike {\bf NA}-querying, we adaptively and sequentially select the next node based on the respondent's answer in a given budget, where the policy of choosing the next node becomes critical. In the following subsection, we will describe complexity to handle all the policies for selecting the next respondent and then introduce the algorithm class of our interest.

\subsection{Challenges and Algorithm Class}
\label{subsec:algorithm class AD}
To see the technical challenges of obtaining an optimal algorithm in the entire algorithm set of {\bf AD}-querying more precisely, we first denote by a vector $Z_{r,i} := Z_{r,i}(p,q) := (x_i (p),Y_i (q))$ where
$x_i (p)$ and $Y_i (q)$ are the answers for the id/dir query which are already defined in Section  \ref{subsec:algorithm class NA} for the respondent $i$.
Let $\mathcal{P}(v_I)$ be a set of all policies, each of which provides
a rule of choosing a next respondent at each querying step, when the
initial respondent is $v_I.$
We denote $W(P)=\{w_1,\ldots,w_{K/r}\}$ as set of selected queried nodes under the
next node selection policy $ P \in \mathcal{P}(v_I)$ with the $i$-th respondent $w_i$ for
$1\leq i \leq K/r$ and denote $A_r (P):= [Z_{r,1},\ldots,Z_{r,K/r}]$ as the answer vector
for all queried nodes under the policy $P$.
Then, we see that it is pretty challenging to find an optimal policy $P$, because
$P$'s action at each $i$-th respondent can be considered as a
mapping $\mathcal{F}_i$ that uses the entire history of the respondents and their answers:
\begin{align}\label{eqn:policy}
\mathcal{F}_i : \{Y_{1}(q),Y_{2}(q),\ldots,Y_{i-1}(q); w_1, \ldots, w_{i-1}\}\rightarrow V_N,
\end{align}
for each $i$. As an approximation, it is natural consider the mapping
$\mathcal{F}_i : (Y_{i-1}(q), w_{i-1}) \rightarrow V_N$, \ie, the next
respondent is determined only by the information at the moment.
it is natural to consider an algorithm based on MLE over all the policies, to maximize
the detection probability, that solves the following optimization:

\separator
\begin{equation*}
 \begin{aligned}
\label{eqn:OPT1}
\text{\bf OPT-AD:} &  \quad \max_{1 \leq r \leq K} \max_{v_I \in V_N}\max_{\substack{ P \in \mathcal{P}(v_I),\\ v \in W(P)}}
\bprob{G_N, A_r (P)|v=v_1, v_I}.
 \end{aligned}
\end{equation*}
\vspace{-0.4cm}
\separator
\vspace{-0.3cm}

 \smallskip
 \noindent{\bf \em Challenges.}
As in the non-adaptive querying, it is important to obtain an analytical
form of the solution of the following problem, to choose the right $r$:
for a fixed $ 1\le r \le K$:
\begin{align}\label{eqn:subOPT1}
\text{\bf SUB-OPT-AD:} &  \quad \max_{v_I \in V_N}\left(\max_{\substack{v \in
                    W(P),\\ P \in \mathcal{P}(v_I)}}
\bprob{G_N, A_r (P)|v=v_1, v_I}\right).
\end{align}

To solve {\bf SUB-OPT-AD}, consider the probability
$\prob{A_r (P)|v=v_1, v_I}$ in \eqref{eqn:subOPT1} for a given $v_I$.
First, it is pretty challenging to find an optimal policy $P$, because
$P$'s action at each $i$-th respondent can be considered as a
mapping $\mathcal{F}_i$ that uses the entire history of the respondents and their answers
as discussed in \cite{Choi17}.
Hence, as an approximation, it is natural consider that the next
respondent is determined only by the information at the moment.
Then, we have the inner part of \eqref{eqn:subOPT1} by
\begin{align}\label{eqn:sequential}
\bprob{&G_N, A_r (P)|v=v_1, v_I}\cr
&=\bprob{G_N|v=v_1, v_I}\times \bprob{A_r (P)|v=v_1, v_I}\cr
&=\bprob{G_N|v=v_1}\times \bprob{Z_{r,K/r},\ldots,Z_{r,1}|v=v_1, v_I}\cr
&=\bprob{G_N|v=v_1}\times \bprob{Z_{r,1}|v=v_1, v_I}\times  \cr
&\qquad \cdots \times \bprob{Z_{r,K/r}|v=v_1, Z_{r,K/r -1}},
\end{align}
where the first equality comes from the fact that the snapshot $G_N$ and the answer data from policy $P$, $A_r (P)$ are independent. The last equality is due to the fact that $Z_{r,i}$ are independent for all $1\leq i \leq K/r$.
Even under
this approximation, this is also not easy to analyze because for a fixed node $v$, the
probability that it is a true parent requires to compute the probability
that the true source is located in $v$'s subtree which does not contain
$w_i$ and there are $O(K(r-1)!)$ different answers for the direction questions.
Thus, we propose a heuristic algorithm that is designed to produce an
approximate solution of {\bf OPT-AD}. The key of our approximate
algorithm is to choose the policy that allows us to analytically compute
the detection probability for a given $r$ so as to compute $r$ easily,
yet its performance is close to that of {\bf OPT-AD}. 
Hence, we restrict our consideration of the following class of AD-querying mechanisms, denoted by $\mathcal{AD}(r,K)$, in this paper:


\smallskip
\begin{definition}(Class $\mathcal{AD}(r,K)$)
\label{def:ad}
In this class of {\bf AD}-querying schemes with the repetition count $r$ and a given budget $K,$
the querier first chooses the RC as a starting node, and performs the repeated procedure mentioned earlier, but in choosing the next respondent, we only consider one of the neighbors of the previous node, where each chosen respondent is asked the id/dir question $r$ times. If the respondent can not obtain any information about the direction (due to all ``yes'' answers for id questions),
it chooses one of the neighbors as the next respondent uniformly at random.
\end{definition}

\smallskip
 From the defined algorithm class $\mathcal{AD}(r,K)$, we  obtain the necessary and sufficient budgets which achieve the target detection probability as follows.


\smallskip
 \subsection{Necessary Budget}
\label{subsec:necessaryAD} 
The necessary budget, which is an information theoretic lower bound for the target detection probability $1-\delta$ for the algorithms in the class $\mathcal{AD}(r,K),$ is presented in the following  Theorem~\ref{theorem:lowerad} by choosing $r$, appropriately.

\smallskip
\begin{theorem}
\label{theorem:lowerad}
Under $d$-regular tree $G$, as $N \rightarrow \infty,$ for any $0< \delta <1,$ there exists a constant $C=C(d),$ such that if
\begin{align}
\label{eqn:lowerad}
     K & \leq \frac{C \cdot H(\mathcal{T}(r^\star)) (\log(7/\delta))^{\alpha/2}}{f_{3} (d,p,q)\log(\log (7/\delta))},
    \end{align}
    for $\alpha=2$ if $p <1$ and $\alpha=1$ if $p=1$ where
\begin{align}
&f_{3} (d,p,q) =(1-H(p))+p(\log_2 d -H(q)), \cr
& r^\star  =  \left \lfloor 1+\frac{7 dp\{3H(p)+2dH(q)\}\log \log K}{2(d-1)}\right\rfloor,
\end{align}
   then no algorithm in the class $\mathcal{AD}(r,K)$ can achieve the detection
probability $1-\delta.$
\end{theorem}

\smallskip
We describe the implications of Theorem~\ref{theorem:lowerad} as follows.
First, when $p$ goes to $1/2$ and $q$ goes to $1/d$, \ie, no information from the querying causes diverging the required budget (because $f_{3} (d,p,q)$ becomes zero). Second, the positive untruthfulness for the id question ($p<1$) requires $\log^{1/2} (1/\delta)$ times more budget than that under the perfect truthfulness ($p=1$). This is because more sampling is necessary to learn the source from the answers of the id questions when $p <1$, whereas no such learning is required for finding the source when $p=1$. Third, large truthfulness (\ie, large $p$) gives more chances to get the direction answers which decreases the amount of budget. Finally, we see that the order is reduced from $1/\delta$ to $\log(1/\delta)$, compared to that in Theorem~\ref{theorem:lower}.

\begin{algorithm}[t!]
{
 \KwIn{Diffusion snapshot $G_N$, querying budget $K$, degree $d$, truthful
   probabilities $p>1/2$, $q>1/d$}
   \KwOut{Estimated rumor source $\hat{v}$ \par\noindent\hrulefill}

\smallskip
$S_I = S_D = \emptyset$ and $\eta (v)= 0$ for all $v \in V_N$\;
Set the initial node $s$ by RC\;
\While{$K\geq r$}{
    {\If{$p=1$}{ If $s=v_1$, return $\hat{v}=s$ otherwise, go to step 2\;}\Else{
    \textbf{Step1}: Set $\eta (s) \leftarrow \eta (s)+1$ which describes that the node $s$ is taken as a respondent and count the number of ``yes''es for the
identity question, stored at $\mu(s)$, and if $\mu(v)/r \geq 1/2$ then
add $v$ to $S_I$\;
}
\smallskip
   \textbf{Step2}: Count the number of
       ``designations'' for the
       direction question among $s$'s neighbors, and choose the largest
       counted node as the predecessor with a random tie breaking\;
       Set such chosen node by $s$ and $K \leftarrow K-r$\;
      }  { } }
$S_D \leftarrow \arg\max_{v\in V_N}
   \eta (v)$\;
    \If{$S_I \cap S_D=\emptyset$}{ $ \hat{v} \leftarrow \arg\max_{v\in S_I \cup S_D}
   \mathbb{P}(G_N | v=v_1)$\; } \Else{ $ \hat{v} \leftarrow \arg\max_{v\in S_I \cap S_D}
   \mathbb{P}(G_N | v=v_1)$\; }
 Return $\hat{v}=s$\;
}
\caption{{\bf MVAD}$(r)$}
\label{alg:interactive}
\end{algorithm}


\smallskip
 \subsection{Sufficient Budget}
\label{subsec:sufficientAD} 
In {\bf AD}-querying, we also consider a simple estimation algorithm
to obtain a sufficient budget named by {\bf MVAD}$(r),$ which is again based on majority voting for both the id and dir questions.
In this algorithm, we choose the RC as the initial node and perform different querying procedures for the following two cases: (i) $p=1$ and (ii) $p<1$. First, when $p=1$, since there is no untruthfulness of the answers of the id questions, we check whether the current respondent $s$ is the source or not. If yes, then the algorithm is terminated and it outputs the node $s$ as a result (Line 5). If not, it asks of $s$ the dir question $r$ times and chooses one predecessor by majority voting with random tie breaking (Line 8). Then, for the chosen respondent, we perform the same procedure until we meet the source or the budget is exhausted. Second, when $p <1,$ we first add one in $\eta (s)$, which is the count that the node $s$ is taken as the respondent. Next, due to untruthfulness, we count the number of ``yes'' answers for the id question and apply majority voting to filter out the nodes that are highly likely to be the source and save them in $S_I$ (Line 7). For the negative answers for id questions, we count the designations of neighbors and apply majority voting to choose the next respondent. Then, we perform the same procedure to the chosen node and repeat this until the budget is exhausted. To filter out more probable source node from the direction answers, we compare the number that is taken as the respondent by designation from the neighbors in $\eta (v),$ and we choose the node which has the maximal count of it and save them into $S_D$ (Line 10). Finally, we
select a node with maximal likelihood in $S_I \cap S_D$ or $S_I \cup S_D$ (Lines 11-14). We easily see that the time complexity of this algorithm is $O(\max\{N,K\})$.

\smallskip
\noindent{\bf \em Rationale:} We now provide the rationale of {\bf MVAD}$(r)$ from the perspective of how we handle the
analytical challenges in \eqref{eqn:sequential}  so as to solve
{\bf MVAD}$(r)$ in an approximate manner.
First, for the identity questions, the intuition is
    similar to the {\bf MVNA}$(r)$ because it uses simple MV-based rule
    for filtration. However, for the direction
    questions with the next respondent selection, we see that a node that has been many designated has large value $\prob{A_r (P)|v=v_1, v_I}$ because 
   \begin{align}
&\prob{Z_{r,i}(p,q)|v=v_1, Z_{r,i -1}}\cr
&=\prob{x_{i}(p)|v=v_1, Z_{r,i -1}}\times \prob{Y_{i}(q)|v=v_1, Z_{r,i -1}},
\end{align}
    for the node $i$
    in \eqref{eqn:sequential}. Hence, the probability $\prob{Y_{i}(q)|v=v_1, Z_{r,i -1}}$ increases.

\smallskip
In selecting a parent node of the target respondent, instead of the exact
calculation of MLE, a simple majority voting is used by selecting the
node with the highest number of designations, motivated by the fact that
when $q>1/d$, such designation sample can provide a good clue of who is
the true parent.

Now, Theorem~\ref{theorem:interactive} quantifies the sufficient amount of budget
to obtain arbitrary detection probability by
appropriately choosing the number of questions to be asked.

\smallskip
\begin{theorem}
\label{theorem:interactive}
For any $0< \delta <1,$ the detection
probability under $d$-regular
tree $G$ is at least $1-\delta,$ as $N \rightarrow \infty,$ if
\begin{align}
\label{eqn:elower1}
    K\geq \frac{2(2d-3)/d(\log(7/\delta))^{\alpha}}{f_{4} (d,p,q)\log (\log (7/\delta))},
    \end{align}
   where $f_{4} (d,p,q)=\frac{2d}{d-1} (p-1/2)^2 + \frac{d-1}{d-2} (q-1/d)^3$ and
     $\alpha=2$ if $p <1$ and $\alpha=1$ if $p=1$ under {\bf MVAD}$(r^\star)$, where
       \begin{align*}
r^\star &= \left\lfloor1+\frac{7d^2 \{2(1-p)^3 + (1-q)^2\} \log\log
  K}{3(d-1)}\right\rfloor.
\end{align*}
\end{theorem}

We see that the gap between necessary and sufficient budgets is $\log(1/\delta)$ when $p<1,$ and $\log^{1/2} (1/\delta),$ when $p=1.$\footnote{The result for $p=1$ matches to the theorem 2 in \cite{Choi17}.}
Note that we have $\log(1/\delta)$ factor reduction from what is sufficient under {\bf MVNA}$(r^\star)$ in the non-adaptive case.
Further, as expected, we see that the sufficient
budget arbitrarily grows as $p$ goes to $1/2$ and $q$ goes to $1/d$, respectively.

\subsection{Adaptivity Gap: Lower and Upper Bounds}
\label{subsec:adaptive}
Using our analytical results stated in Theorems~\ref{theorem:lower}-\ref{theorem:interactive}, we now establish the quantified adaptivity gap defined as follows:

\smallskip
\begin{definition}(Adaptivity Gap)
Let $K_{na}(\delta)$ and $K_{ad}(\delta)$ be the amount of budget needed to obtain $(1-\delta)$ detection probability for $0< \delta <1$ by the optimal algorithms in the classes $\mathcal{NA}(r,K)$ and $\mathcal{AD}(r,K),$ respectively. Then, the adaptivity gap, $\text{AG}(\delta)$ is defined as $K_{na}(\delta)/K_{ad}(\delta).$
\end{definition}
\smallskip


\smallskip
\begin{theorem}
\label{thm:AG}
For a given $0<\delta<1,$ there exist a constant $r$ and two other constants $U_1 = U_1(r,p,q)$ and $U_2 = U_2(r,p,q)$, where
the constant $r$ corresponds to the number of repeated id/dir questions for each respondent in both classes $\mathcal{NA}(r,K)$ and $\mathcal{AD}(r,K)$, such that
\begin{align}
\label{eqn:elower2}
   \frac{U_1 \cdot (1/\delta)^{1/2}}{\log^{\alpha} (1/\delta)} \leq AG(\delta) \leq \frac{U_2 \cdot (1/\delta)}{\log^{\alpha/2} (1/\delta)},
    \end{align}
   where $\alpha=2$ if $p <1$, and $\alpha=1$ if $p=1$.

\end{theorem}

\smallskip
In Theorem~\ref{thm:AG}, we see that
for a given target detection probability $1-\delta$,
the amount of querying budget by adaptive querying asymptotically decreases at least from $(1/\delta)^{1/2}$ to $\log^{2}(1/\delta).$
This implies that there is a significant gain of querying
in the adaptive manner.
Further, the difference of upper and lower bounds of $\text{AG}(\delta)$ is expressed by square root in our algorithm classes, when we use {\bf MVNA}$(r^\star)$ and {\bf MVAD}$(r^\star)$ for sufficient budgets, respectively.

\section{Proofs}\label{sec:proof}
In this section, we will provide the
proof sketches for the Theorems due to the page limit. The whole proof will be provided in our supplementary material \cite{Jae16}.

\subsection{Proof of Theorem~\ref{theorem:lower}}
For a given $r,$ we introduce the notation $V_l,$ which is
equivalent to $C_r,$ where the hop distance $l=\frac{\log\left(\frac{K(d-2)}{rd}+2\right)}{\log(d-1)}$.
 Also for notational simplicity, we simply use $
\prob{\hat{v}=v_{1}}$ to refer to $\lim_{N \to \infty }\prob{\hat{v}(G_N,r)=v_{1}}$
for any estimator given the snapshot $G_N$ and redundancy parameter $r$ in the proof section.
Then, the detection probability is expressed as the product of the two
terms:
    \begin{align}
      \label{eqn:detect0}
      \prob{\hat{v} = v_{1}} &=   \prob{v_1 \in  V_{l}}\times  \prob{\hat{v}=v_1|v_1 \in V_{l}},
    \end{align}
where the first one is the probability that the source is in the $l$-hop based candidate set $V_l$
and the second term is the probability that the estimated node is exactly the source in the candidate set for
any learning algorithm under the algorithm class $\mathcal{C}(l,r)$.
We first obtain the upper bound of probability of first term in \eqref{eqn:detect0}
in the following lemma.

\smallskip
\begin{lemma}\label{lem:multihop_upper}
For $d$-regular trees,
\begin{equation}
\prob{v_1 \in V_{l}} \leq 1-c\cdot e^{-l\log l},
\label{eqn:multihop_upper}
\end{equation}
where $c = (d-2)/4d$.
\end{lemma}
\smallskip

We will closely look at the case of each $l$, to derive the
probability that the rumor center $v_{RC}$ is exactly $l$-hop distant
from the rumor source $v_1$. Let $\delta_1$ be the error for the $\prob{v_1 \notin V_{l}}$ then
it is lower bounded by $\delta_1 \geq c\cdot e^{-l\log l}$.

To obtain the second term in \eqref{eqn:detect0}, we use the information theoretical techniques
for the direct graph inference as done in \cite{Sujay2012} with partial observation because, if the rumor spread from the source
we can obtain a direct tree where all direction of edges are outgoing from the source.
From the assumption
of independent answers of queries, we see that the snapshot from one querying process with untruthful for direction question is equivalent to the
snapshot of diffusion flow from the source under the IC-diffusion model with noisy observation.
By using these fact and the result of graph learning techniques from the epidemic cascades in \cite{Sujay2012}, we obtain the following lemma.

\smallskip
\begin{lemma}\label{lem:information}
For any graph estimator to have a probability of error of $\delta_2>0$, it needs $r$ queries to the candidate
set $V_l$ with $|V_l|=n$ that satisfies
\begin{equation}\label{eqn:information1}
r\geq \frac{\log(1/\delta_2)H(T) (n-1)\log \frac{n}{2}}{ n((1-H(p))+p (1-p)(\log_2 d-H(q)))},
\end{equation}
where $H(T)$ is the entropy of infection time vector and $H(p)=p \log p + (1-p)\log (1-p)$ and $H(q)= q \log q + (1-q)\log \frac{1-q}{d-1} $, respectively.
\end{lemma}
\smallskip

This result indicates that if there is no information from query, \ie, $p=1/2$ and $q=1/d$, the required number of queries diverges. Further, if the uncertainty of infection time $H(T)$ for the nodes in $V_l$ increases, the required queries also increases. Then, from the disjoint of two error event and by setting $\delta_1 = \delta_2 =\delta /2$ with $l=\log\left(\frac{K(d-2)}{rd}+2\right)/\log(d-1)$, we have
  \begin{align*}
     \prob{&\hat{v} \neq v_{1}} \geq c\cdot e^{-\frac{\log\left(\frac{K}{r}\right)}{\log(d-1)}\log \frac{\log\left(\frac{K}{r}\right)}{\log(d-1)}} \cr
     & +
     e^{-\frac{H(T)(\frac{K}{r}-1)\log \frac{K}{2r}}{K((1-H(p))+p (1-p)(\log_2 d-H(q)))}}
     \geq \delta.
    \end{align*}
From the fact that $\lambda =1$ in our setting and Lemma 2 in \cite{Sujay2012}, we obtain $H(T)\leq K/r $ and by differentiation of above lower bound with respect to $r$, we approximately obtain
$r^\star= \left \lfloor 1+\frac{4 (1-p)\{7H(p)+H(q)\}\log K}{3e\log(d-1)}\right\rfloor$ where
the derivation is given in the supplementary material.
Since if we use the $r^\star$, it gives the upper bound of detection probability hence,
we put it to the obtained upper-bound which is expressed as a
function of $K,$ as follows:
\begin{align}
  \label{eq:kkk1}
  &\prob{\hat{v} \neq v_{1}}  \cr
  & \geq  \frac{1}{2}\ e^{-h_1 (T,p,q) \log K\log (\log K)}+\frac{c}{4} e^{- 2 h_1 (T,p,q)\log K\log (\log K)}\cr
&   \geq  C_d e^{-2 h_1 (T,p,q)\log K \log (\log K)},
\end{align}
where $C_d=(c+3)/4$ and $h_1 (T,p,q)= H(T)^{-1}(1-H(p))+p(1-p)(\log_2 d-H(q))$. If we set $\delta \leq C_d e^{-2 h_1 (T,p,q)\log K\log (\log K)},$ we find the value $K$ such that its assignment to
\eqref{eq:kkk1} produces the error probability $\delta,$ and we finally obtain the
desired lower-bound of $K$ as in Theorem~\ref{theorem:lower}.

\subsection{Proof of Theorem~\ref{theorem:noninteractive}}

We first provide the lower bound on detection probability of {\bf
  MVNA}$(r)$ for a given $K$ and $r$ in the following lemma.

\smallskip
\begin{lemma}
  \label{lem:simple_prob}
  For $d$-regular trees ($d\geq 3$), a given budget
  $K,$ our estimator $\hat{v}$ from {\bf MVNA}$(r)$ has the
  following lower-bound of the detection probability:
\begin{multline}
  \label{eqn:detect}
\prob{\hat{v}=v_{1}}\geq 1-c\left(\frac{r+p+q}{r+2}\right)^3\cdot
  \exp\Bigg( \frac{-h_d
  (K,r)w_d (p,q)}{2}\Bigg),
\end{multline}
where $c = 7(d+1)/d$ and $w_d (p,q)=\frac{1}{2}(4(p-1/2)^2+(d/(d-1))^3 (q-1/d)^3)$.
The term $h_d (K,r)$ is given by
$$h_d (K,r):=\frac{\log\left(\frac{K}{r}\right)}{\log(d-1)}\log\left(
\frac{\log\left(\frac{K}{r}\right)}{\log(d-1)}\right).$$
\end{lemma}
\smallskip

\begin{proof}
Under the {\bf MVNA}$(r)$, the detection probability is expressed as the product of the three
terms:
    \begin{align}
      \label{eqn:detection}
      \prob{\hat{v} = v_{1}} &=   \prob{v_1 \in  V_{l}}\times  \prob{\hat{v}=v_1|v_1 \in V_{l}} \cr
                               & = \prob{v_1 \in  V_{l}} \times \prob{v_1
                                 \in \hat{V}|v_1 \in V_{l}}   \cr
                               &  \times \prob{v_1 =v_{LRC}|v_1 \in \hat{V}},
    \end{align}
where $\hat{V}:=S_I \cap S_D$ if it is not empty or $\hat{V}:=S_I \cup S_D$, otherwise. This is the filtered candidate set in {\bf
  MVNA}$(r)$ and  $v_{LRC}$ is the node in $\hat{V}$ that has the
highest rumor centrality \ie, likelihood, where $LRC$ means the local rumor center.
We will drive the lower bounds of the first, second, and the third terms
of RHS of \eqref{eqn:detection}.
The first term of RHS of \eqref{eqn:detection} is bounded by
\begin{align}
  \label{eq:first}
  \prob{v_1 \in V_{l}} \geq 1-c\cdot e^{-(l/2)\log l},
\end{align}
where the constant $c = 7(d+1)/d$ from Corollary 2 of \cite{Khim14}.
Let $S_N$ be the set of revealed nodes itself as the rumor source and let $S_I$ be the set of nodes which minimizing the errors.
If the true source is in $V_{l},$ then the probability that it is most indicated node for
a given budget $K$ with the repetition count $r$ and truth probability $p>1/2$ and $q>1/d$ is given by
    \begin{align}\label{eqn:d}
      \prob{v_1 &=v_{LRC}|v_1 \in \hat{V}}\cr
     & =\prob{v_1= \arg\max_{v\in S_I\cap S_D}R(v,G_{N})|K,p,q}.
    \end{align}
To obtain this, we consider that
if $p>1/2$,
 the probability $v_1 \in S_I$ by the majority voting, because the selected node can be designation again
in the algorithm. We let total number of queries by $r\geq1$, we let $W=\sum_{i=1}^{r}X_i (v_1)$ for the source node $v_1$, then
the probability that true source is in the filtration set $S_I$ is given by $\prob{W\geq r/2}=\sum_{j=0}^{\lfloor r/2\rfloor}\binom{ r}{j}(1-p)^{j}p^{r-j}.$
Then, from this relation, we have the following lemmas
whose proofs are will be provided in \cite{Choi17}:
\smallskip
\begin{lemma}(\cite{Choi17})
When $p>1/2,$
\begin{eqnarray*}
\prob{v_1 \in S_I|v_1 \in V_{l}}  & \geq &p+(1-p)(1-e^{-(p-1/2)^{2}\log r}).
\end{eqnarray*}
\label{lem:majority}
\end{lemma}
This result implies the lower bound of probability that the
source is in $S_I$ for a given $r$. Next, we will obtain the probability that the source is in $S_D$ after filtration of the direction answers.
To do this, we first consider that the total number of direction queries $N_d$ is a random variable which is given by:
\begin{align*}
P(N_d = k)=\begin{cases}
\binom{r}{k}p^{r -k}(1-p)^{k}&\mbox{if}~v=v_1\\
\binom{r}{k}(1-p)^{r -k}p^{k}&\mbox{if}~v \neq v_1,
\end{cases}
\end{align*}
where $k$ is less than parameter $r$. Using this fact, we obtain the following result.
\smallskip

\begin{lemma}
When $p>1/2$ and $q>1/d$,
\begin{eqnarray*}
\prob{v_1 \in S_D|v_1 \in V_{l}}  & \geq & 1- e^{-\frac{rp(d-1)(q-1/d)^2}{3d}}.
\end{eqnarray*}
\label{lem:consistence}
\end{lemma}
\smallskip

This result shows the lower bound of probability that the source is in $S_I$ for a given $r$.
By considering the two results in the above, we have the following lemma.

\smallskip
\begin{lemma}
For given repetition count $r$, we have
\begin{equation}\label{eqn:d1}
    \begin{aligned}
      P(v_1 \in S_I \cap S_D|v_1 \in V_l) \geq 1-2e^{-f(p,q) 2r\log r}
    \end{aligned}
  \end{equation}
where $f (p,q)=3(p-1/2)^2 + \frac{d-1}{3d} p(1-p)(q-1/d)^2$.
\label{lem:finalfil}
\end{lemma}
\smallskip

Then, we obtain the following lemma, which is the lower bound of detection probability
among the final candidate set.

\smallskip
\begin{lemma} \label{lem:filtration}
  When $d\geq3$, $p>1/2$ and $q>1/d$,
  \begin{eqnarray*}
    \prob{v_1 =v_{LRC}|v_1 \in S_I \cap S_D}  & \geq & 1-e^{-f(p,q) r\log r}.
  \end{eqnarray*}
\end{lemma}
\smallskip

Merging these lower-bound with the lower-bound in
\eqref{eq:first} where we plug in
$l=\frac{\log\left(\frac{K(d-2)}{rd}+2\right)}{\log(d-1)},$ we finally get the lower bound of detection probability
 of {\bf MVNA}$(r)$ for a given repetition count $r$ and this completes the proof of
 Lemma~\ref{lem:simple_prob}.
\end{proof}

To finish the proof of theorem, note that the second term of RHS of \eqref{eqn:detect} is the probability that the
source is in the candidate set for given $K$ and $r$. Hence, one can see
that for a fixed $K$, large $r$ leads to the decreasing detection
probability due to the smaller candidate set.  However, increasing $r$
positively affects the first term of RHS of \eqref{eqn:detect}, so that
there is a trade off in selecting a proper $r$.
By derivation of the result with respect to $r$, we first obtain $r^\star$ which maximizes the detection
probability by $r^\star= \left \lfloor 1+\frac{2(1-p)\{1+(1-q)^2\}\log K}{e\log(d-1)}\right\rfloor$ in {\bf MVNA}$(r^\star)$
and put this into the error probability $\prob{\hat{v}\neq v_{1}}$ such as

\begin{align}
  \label{eq:ppp}
  \prob{\hat{v}&\neq v_{1}} \leq e^{-f(p,q) r\log r}+2e^{-f(p,q) 2r\log r}+c\cdot e^{- \frac{l}{2}\log l},
\end{align}
where the constant $c$ is the same as that in \eqref{eq:first}.
Now, we first put
$l=\frac{\log\left(\frac{K(d-2)}{rd}+2\right)}{\log(d-1)}$ into
\eqref{eq:ppp} and obtained the upper-bound of \eqref{eq:ppp}, expressed
as a function of $r,$ for a given $p$ and $q$ and the constant $c.$
Then, we take $r^*$ and put it to the obtained upper-bound which is expressed as a
function of $K,$ as follows:
\begin{align}
  \label{eq:kkk2}
  \prob{\hat{v} \neq v_{1}}  & \leq  3\ e^{-f(p,q) \log K\log (\log K)}+c e^{- \frac{\log K}{2}\log (\log K)}\cr
&   \leq  c_1 e^{-f (p,q)\frac{\log K}{2} \log (\log K)},
\end{align}
where $c_1=c+3$. If we set $\delta \geq c_1 e^{-f (p,q)\frac{\log K}{2}\log (\log K)},$ we find the value of $K$ such that its assignment to
\eqref{eq:kkk2} produces the error probability $\delta,$ and we get the
desired lower-bound of $K$ as in the theorem statement. This completes
the proof of Theorem~\ref{theorem:noninteractive}.

\subsection{Proof of Theorem~\ref{theorem:lowerad}}
We will show the lower bound for given $K$ and $r$ of the case $p<1$. \footnote{The result
  for $p=1$ is similar to this except the termination of querying process when it meets the source.}
For a given $r,$ we let $V_L$ be the set of all infected nodes from the rumor center within a distance $L:=K/r$ then we see that
the querying dynamic still becomes a directed tree construction rooted by the source $v_1$.
Then, the detection probability is expressed as the product of the two
terms:
    \begin{align}
      \label{eqn:detect01}
      \prob{\hat{v} = v_{1}} &=   \prob{v_1 \in  V_{L}}\times  \prob{\hat{v}=v_1|v_1 \in V_{L}},
    \end{align}
where the first one is the probability that the distance between source and rumor center is less than $K/r$
and the second term is the probability that the estimated node is exactly the source in the candidate set for
any learning algorithm under the algorithm class $\mathcal{AD}(r,K)$.
First, from Lemma~\ref{lem:multihop_upper}, we have that the probability of first term in \eqref{eqn:detect01} is
upper bounded by $1-ce^{-(K/r)\log(K/r)}$ where $c=4d/3(d-2)$ for a given budget $K$ and repetition count $r$.
We see that
the querying dynamic still becomes a directed tree construction rooted by the source $v_1$.
However, different to the NA-querying, the querying process gives direction data of a subgraph of the original direct tree
because the querier chooses a node, interactively.
For a given $r$, let $Z_{r,i}$ be the answer data of querying for a selected queried node $i$ where $1 \leq i \leq K/r.$ Then,
from the assumption of the algorithm class $\mathcal{AD}(r,K)$,
the joint entropy for the random answers with the infection time random vector $T$,
$H(T, Z_{r,1},\ldots,Z_{r,K/r})$ is given by
 \begin{align}\label{eqn:entropy}
     H(T, Z_{r,1},&\ldots,Z_{r,K/r})=\sum_{i=1}^{K/r}H(T, Z_{r,i}|Z_{r,i-1},\ldots,Z_{r,1})\cr
     &=\sum_{i=1}^{K/r}H(T, Z_{r,i}|Z_{r,i-1})\stackrel{(a)}{=}\sum_{i=1}^{K/r}H(T, Z_{r,i}),
    \end{align}
where $(a)$ is from the fact that all data $Z_{r,i}$ are independent.
Let $G^*$ be the true directed graph and let $\hat{G}$ be be an estimated directed tree from the
sequential answers of adaptive querying $(Z_{r,1},\ldots,Z_{r,K/r})$. Then, we see that
this defines a Markov chain
$$G^* \rightarrow (T, Z_{r,1},\ldots,Z_{r,K/r}) \rightarrow \hat{G},$$
from the defined algorithm class $\mathcal{AD}(r,K)$. By property of the mutual information, we have
 \begin{align}\label{eqn:entropy1}
    I(G^*; &T, Z_{r,1},\ldots,Z_{r,K/r})\cr
    &\leq H(T, Z_{r,1},\ldots,Z_{r,K/r})=\sum_{i=1}^{K/r}H(T, Z_{r,i}) \cr
    &\stackrel{(a)}{=} (K/r)H(T, Z_{r,1})\cr
    &\stackrel{(b)}{\leq} (KH(T)/r)[r(1-H(p))+rp(\log_2 d - H(q))]\cr
    &= KH(T)[(1-H(p))+p(\log_2 d - H(q))]\cr
    &:=Kh(p,q),
    \end{align}
where $(a)$ follows from the fact that the answers $Z_{r,i}$ are mutually exclusive
and $(b)$ is from the fact that $H(T, Z_{r,1})=(1-H(p)+rp(\log_2 d - H(q)))/H(T)$ since
the number of direction answers follows binomial distribution. Let $\mathcal{G}_{K/r}$ be the
set of possible directed tree in $V_s$ then we have $|\mathcal{G}_{K/r}|\leq (K/r)\log (K/2r)$.
Using the Fano's inequality on the Markov chain $G^* \rightarrow (Z_{r,1},\ldots,Z_{r,K/r}) \rightarrow \hat{G},$
we obtain
 \begin{align}\label{eqn:entropy2}
   \prob{G \neq G^*}
   &\geq  \frac{I(G^*; Z_{r,1},\ldots,Z_{r,K/r})+h(p,q)}{H(T)\log|\mathcal{G}_{K/r}|}\cr
   & \geq  \frac{Kh(p,q)+h(p,q)}{\frac{KH(T)}{r}\log (\frac{K}{2r}-1)}.
    \end{align}
From the disjoint of two error event and by setting $\delta_1 = \delta_2 =\delta /2$ for each error, we have
  \begin{align}\label{eqn:entropy3}
     \prob{&\hat{v} \neq v_{1}} \geq c\cdot e^{-(K/r)\log(K/r)} \cr
     & +e^{-\frac{\frac{KH(T)}{r}\log (\frac{K}{2r}-1)}{Kh(p,q)+h(p,q)}}
     \geq \delta.
    \end{align}
From the fact that $\lambda =1$ in our setting and Lemma 2 in \cite{Sujay2012}, we obtain $H(T)\leq K/r $ and
by differentiation of above lower bound with respect to $r$, we approximately obtain
$r^\star= \left \lfloor 1+\frac{7 dp\{3H(p)+2dH(q)\}\log \log K}{2(d-1)}\right\rfloor$ where
the derivation is given in the supplementary material.
Since if we use the $r^\star$, it gives the upper bound of detection probability hence,
we put it to the obtained upper-bound which is expressed as a
function of $K,$ as follows:
\begin{align}
  \label{eq:kkk3}
  \prob{\hat{v} \neq v_{1}}  & \geq  \frac{1}{3}\ e^{-h_2(T,p,q) K\log (\log K)}+\frac{c}{4} e^{- 7 h_2 (T,p,q)K\log (\log K)}\cr
&   \geq  C_d e^{-7h_2 (T,p,q) K\log (\log K)},
\end{align}
where $C_d=2(c+3)/7$ and $h_2 (T,p,q)= H(T)^{-1}(1-H(p))+(1-p)(\log_2 d-H(q))$. If we set $\delta \leq C_d e^{-7 h_2 (T,p,q)K\log (\log K)},$ we find the value of $K$ such that its assignment to
\eqref{eq:kkk3} produces the error probability $\delta,$ and we get the
desired lower-bound of $K$ as in the theorem statement.
Then, we finally obtain the result and this completes the proof of
Theorem~\ref{theorem:lowerad}.

\subsection{Proof of Theorem~\ref{theorem:interactive}}
We will show the lower bound on the detection
  probability for given $K$ and $r$ of the case $p<1$ \footnote{The result
  for $p=1$ is given in \cite{Choi17} and we omit it here.} in Lemma~\ref{lem:inter_prob}.

\smallskip
\begin{lemma}
  \label{lem:inter_prob}
  For $d$-regular trees ($d\geq 3$), a given budget
  $K,$ our estimator $\hat{v}$ from {\bf MVAD}$(r)$ has the
  detection probability lower-bounded by:
\begin{align}
  \label{eqn:detect2}
\prob{\hat{v}=v_{1}} \ge & 1-c (g_d (r,q))^3 \cr
&\cdot \exp\left [-\left(p-\frac{1}{2}\right)^2\left(\frac{K}{r}\right)\log \left(\frac{K}{r}\right) \right],
\end{align}
where $g_d (r,q):= e^{-\frac{r(d-1)(q-1/d)^2}{3d(1-q)}}$ and
$c=(5d+1)/d$.
\end{lemma}
\smallskip

\begin{proof}
For the {\bf MVAD}$(r)$, for a given $r,$ we introduce the notation $V_s,$
where the set of all queried nodes of the algorithm. From the initial queried node, we need the probability
that the source is in the set of queried node by some policy $P \in \mathcal{P}(v_I)$.
Then, the detection probability is also expressed by the product of the three
terms:
    \begin{align}
      \label{eqn:detection1}
      \prob{\hat{v} = v_{1}} &=   \prob{v_1 \in  V_{L}}\times  \prob{\hat{v}=v_1|v_1 \in V_{L}} \cr
                               & = \prob{v_1 \in  V_{L}} \times \prob{v_1
                                 \in \hat{V}|v_1 \in V_{L}}   \cr
                               &  \times \prob{v_1 =v_{LRC}|v_1 \in \hat{V}},
    \end{align}
where $V_{L} = \{v| d(v_{RC},v) \le K/r\}$ because the number of budget is $K$ and $\hat{V}=S_I \cap S_D$
if it is not empty or $\hat{V}=S_I \cup S_D$, otherwise.
From the result in Corollary 2 of \cite{Khim14}, we have $\prob{E_1}\leq c \cdot e^{- (K/r)\log K/r}$
since we use additional direction query with identity question. For the second part of probability in \eqref{eqn:detection1},
we obtain the following lemma.

\smallskip
\begin{lemma}
When $p>1/2,$
\begin{equation*}
 \begin{aligned}
\prob{v_1 &\in S_I|v_1 \in V_{L}}  \\
&\geq \left(p+(1-p)(1-e^{-(p-1/2)^{2}\log r})\right) \left(1-ce^{-\frac{Kp}{r}  (q-1/d)^3}\right).
   \end{aligned}
   \end{equation*}
\label{lem:idinter}
\end{lemma}
\smallskip

\begin{proof}
Let $Q_K (v)$ be the number of queries to a node $v \in V_l$ when there are $K$ queries then we have
\begin{equation*}
    \begin{aligned}
     &\prob{Q_K (v_1)\geq1}\\
     &=\sum_{i=1}^{l}\prob{Q_K (v_1)\geq1|d(v_1, v_{RC})=i}\prob{d(v_1, v_{RC})=i}.
    \end{aligned}
  \end{equation*}
where $\prob{d(v_1,v_{RC}) = i}$ is the probability that the distance from the rumor center to rumor source is $i$ and this probability become smaller if the distance between rumor source and rumor center is larger.
From this, we have the following result for the lower bound of the probability of distance between the rumor center and source.
\smallskip
\begin{proposition}
For $d$-regular trees,
\begin{equation}
\prob{d(v_1,v_{RC}) = i}\geq \left(\frac{d-1}{d}\right)^{i} e^{-(i+1)}.
\label{eqn:interactive_multihop1}
\end{equation}
\label{pro:distance1}
\end{proposition}
\smallskip

Next, we construct the following Markov chain.
Let $\hat{p}:=\prob{W = r}$ for the identity questions \ie, there is no ``no'' answers for the
identity questions so that the algorithm should chooses one of neighbor nodes uniformly at random
and let $\hat{q}:=\prob{Z_1 (v) > Z_j
(v),~ \forall j}$ for the direction question, respectively.
Different to the case for $p=1$ which the node reveals itself as the rumor source or not
with probability one so that the Markov chain has the absorbing state, in this case, there is no such a state.
To handle this issue, we use
the information that how many times the neighbors indicate a node as its parent and how many times a node reveals itself as the rumor source.
To do that, we consider the case that there is a \emph{token}\footnote{The token keeper is regarded as the current respondent in this model.} from the initial state and
it move to the next state after additional querying follows the answer. Then this probability is the same that after $K/r$-step of Markov chain, and we expect that the rumor source $v_1$ will have the largest chance to keeping this token due to the assumption of biased answer. Let $X_n$ be the state (node) which keep this token at time $n$ where
the state is consist of all node in $V_N$. The initial state is the rumor center such that $X_0 =0$ where $0$ indicates the rumor center. Then there are $(d(d-1)^{K/r}-2)/(d-2)$ states and we can index all the state properly. Let $p^{n}_{k,j}$ be the $n$ step transition probability from the state $k$ to the state $j$.
To obtain this probabilities, we first label an index ordering by counter-clockwise from the rumor center $X_0=0$. Then, we have $P(X_{n+1} =k|X_n =k)=0$ for all $k$ and $n$, respectively. Furthermore, $P(X_{n+1} =j|X_n =k)=0$ for all $d(k,j)>1$ since the token is moved one-hop at one-step ($r$ querying). Then, the transition probability for the node $k$ which is not a leaf node in $V_L$ is as follows.
\begin{align*}
p_{k,j}=\begin{cases}
\frac{\hat{p}}{d}+(1-\hat{p})\frac{1-\hat{q}}{d-1}&\mbox{if}~j\notin nb(k,v_1)\\
\frac{\hat{p}}{d}+(1-\hat{p})\hat{q}&\mbox{if}~j \in nb(k,v_1),
\end{cases}
\end{align*}
where $nb(k,v_1)$ is the set of neighbors of the node $k$ on the path between the node $k$ and $v_1$.
%
From the simple Markov property of querying scheme, if we assume that the source node is an absorbing state then
we obtain for a given budget $K/r\geq l$,
    \begin{align}\label{eqn:transition}
     \prob{&Q_K (v_1)\geq1|d(v_1, v_{RC})=i}\cr
     &=1-\prob{Q_K (v_1)=0|d(v_1, v_{RC})=i}\cr
     &=1-\prob{\sum_{n=0}^{K}I_n (v_1)= 0|d(v_1, v_{RC})=i}\cr
     &=1-\prob{I_n (v_1)= 0,~\text{$\forall i\leq n \leq K/r$}|d(v_1, v_{RC})=i}\cr
     &=1-\prod_{n=i}^{K/r}(1-p^{n}_{0,v})\cr
     &\stackrel{(a)}{\geq}1-\prod_{n=i}^{K/r}(1-p^{K/r}_{0,v})=1-(1-p^{K/r}_{0,v})^{K/r-i}\cr
     &\stackrel{(b)}{\geq}1-e^{-(K/r -i)p^{K/r}_{0,v}},
    \end{align}
where $(a)$ follows from the fact that $p^{n}_{0,v}\geq p^{K/r}_{0,v}$ for all $i\leq n \leq K/r$ and $(b)$ is from the relation of $(1-p)^{K/r} =e^{K/r \log (1-p)}\leq e^{-p(K/r)}$ where we use the inequality $\log (1-x) \leq -x$ for $0\leq x \leq 1$. Note that the transition probability is the case of $d(v_1, v_{RC})=i$. From Lemma, we have
\begin{equation*}
    \begin{aligned}
     \prob{Q_K (v_1)\geq1}&\geq\sum_{i=1}^{\infty}(1-e^{-(K/r -i)p^{K/r}_{0,v}})P(d(v_1, v_{RC})=i)\\
     &\geq\sum_{i=1}^{\infty}(1-e^{-(K/r -i)p^{K/r}_{0,v}})\left(\frac{d-1}{d}\right)^{i} e^{-(i+1)}\\
    &\geq 1-ce^{-\frac{Kp^{K/r}_{0,v}}{r} }\geq 1-ce^{-\frac{Kp}{r}  (q-1/d)^3}.\\
    \end{aligned}
  \end{equation*}
Using this result and Lemma~\ref{lem:majority}, we conclude the result of
Lemma~\ref{lem:idinter} and this completes the proof.
\end{proof}

\smallskip
Next, we have the following result.

\smallskip
\begin{lemma}
When $p>1/2$ and $q>1/d,$
\begin{equation*}
 \begin{aligned}
\prob{v_1 \in S_D |v_1 \in V_{L}} \geq 1- e^{-\frac{Kp (d-1)(q-1/d)^2}{6rd}}
   \end{aligned}
   \end{equation*}
\label{lem:directionInter}
\end{lemma}
\smallskip

Similar to the previous one, to obtain the detection probability, we need to find the probability $P(v_1 \in S_N \cap S_I|v_1 \in V_{L})$.
From this, we consider $r$ repetition count for identity question and $r-X_{r}(v)$ for direction question where $(X_{r}(v)$
be the number of yes answers of the queried node $v$ with probability $p_v$. Hence, we see that the number of repetition count
for the direction questions is also a random variable which follows a binomial distribution with parameter $p_v$.
By considering this, We have the following lemma.

\smallskip
\begin{lemma}
Suppose $v_1 \in V_{L}$ then we have
\begin{equation}\label{eqn:candidate}
    \begin{aligned}
      P(v_1 \in S_I \cap S_D |v_1 \in V_{L})\geq 1-e^{-6g(p,q)^2 (K/r)\log r},
    \end{aligned}
  \end{equation}
where $g (p,q)=\frac{2d}{d-1} (p-1/2)^2 + \frac{d-1}{d-2} (q-1/d)^3$.
\label{lem:filter_Inter}
\end{lemma}
\smallskip

\begin{proof}
Since the events $v_1 \in S_I$ and $v_1 \in S_D$ are independent for a given $v_1 \in V_{L}$,
by using lemma \ref{lem:directionInter} and \ref{lem:directionInter} and some algebra, we have
\begin{equation*}
    \begin{aligned}
      &\prob{v_1 \in S_I \cap S_D |v_1 \in V_{L}}\\
      &\geq \left(p+(1-p)(1-e^{-(p-1/2)^{2}\log r})\right) \left(1-ce^{-\frac{Kp}{r}  (q-1/d)^3}\right)\\
      &\qquad \cdot \left(1- e^{-\frac{K (d-1)(q-1/d)^2}{6rd}}\right)\\
      &\geq  1-e^{-6g(p,q)^2 (K/r)\log r},
    \end{aligned}
  \end{equation*}
where $g (p,q)=c_1 (p-1/2)^2 + c_2 (q-1/d)^3$ for some constants $c_1$ and $c_2$ which are only depends on the degree $d$. This completes the proof of Lemma~\ref{lem:filter_Inter}.
\end{proof}

Next, we consider the following lemma which indicates the lower bound of detection probability
among the final candidate set.

\smallskip
\begin{lemma} \label{lem:filtration1}
  When $d\geq3$, $p>1/2$ and $q>1/d$,
  \begin{eqnarray*}
    \prob{v_1 =v_{LRC}|v_1 \in S_I \cap S_D}  & \geq & 1-e^{-g(p,q) K\log r}.
  \end{eqnarray*}
\end{lemma}
\smallskip

The proof technique is similar to the Lemma \ref{lem:filtration} so we omit it.
Using the obtained lemmas 5-8, we finally get the lower bound of detection probability
 of {\bf MVAD}$(r)$ for a given repetition count $r$ and this completes the proof of
 Lemma~\ref{lem:inter_prob}.
\end{proof}

The term $g_d (r,q)$ in
\eqref{eqn:detect2} is the probability that the
respondent reveals the true parent for given $r$ and $q$.
 Hence, one can see
 that for a fixed $K$, large $r$ leads to the increasing this
 probability due to the improvement for the quality of the direction answer.
 However, increasing $r$
 negatively affects the term $K/(r+1)$ in \eqref{eqn:detect2}, so that
 there is a trade off in selecting a proper $r$.
By considering the error probabilities, we obtain
\begin{equation}\label{eqn:error}
\begin{aligned}
\prob{\hat{v}\neq v_{1}}&\leq
c \cdot e^{- (K/r)\log (K/r)}+e^{-3g(p,q)^2 (K/r)\log r}\\
&+e^{-g(p,q) K \log r}\\
&\stackrel{(a)}{\leq}( c
+1)e^{-2g(p,q)^2 (K/r)(\log K/r)},
\end{aligned}
\end{equation}
where $c_1=c+1$ and $g (p,q)=\frac{2d}{d-1} (p-1/2)^2 + \frac{d-1}{d-2} (q-1/d)^3$. The inequality $(a)$ is from the fact that
$g(p,q) <1$.
By derivation of the result with respect to $r$, we first obtain $r^\star$ which maximizes the detection
probability by $r^\star= \left\lfloor1+\frac{7d^2 \{2(1-p)^3 + (1-q)^2\} \log\log
  K}{3(d-1)}\right\rfloor$ in {\bf MVAD}$(r^\star)$
and put this into the error probability $\prob{\hat{v}\neq v_{1}}$, we have
\begin{align}\label{eqn:choi}
\prob{\hat{v}\neq v_{1}}&\leq
( c
+1)e^{-2g(p,q)^2
  (K/(r^*))\log(K/(r^* ))}\cr
   &\stackrel{(a)}{\leq} ( c
+1) e^{-g (p,q)K\log (\log K)},
\end{align}
where the inequality $(a)$
comes from the obtained result of $r^\star$. Let
$\delta \geq ( c
+1) e^{-g(p,q)K\log (\log K)},$ then, we obtain the value of $K$ which
 produces the error probability $\delta$ in later and we obtain the
desired lower-bound of $K$ as in the theorem statement. This completes
the proof of Theorem~\ref{theorem:interactive}.

\begin{figure*}[t!]
\begin{center}
\subfigure[Best $r^*$ with varying $K$.]{\includegraphics[width=0.5\columnwidth]{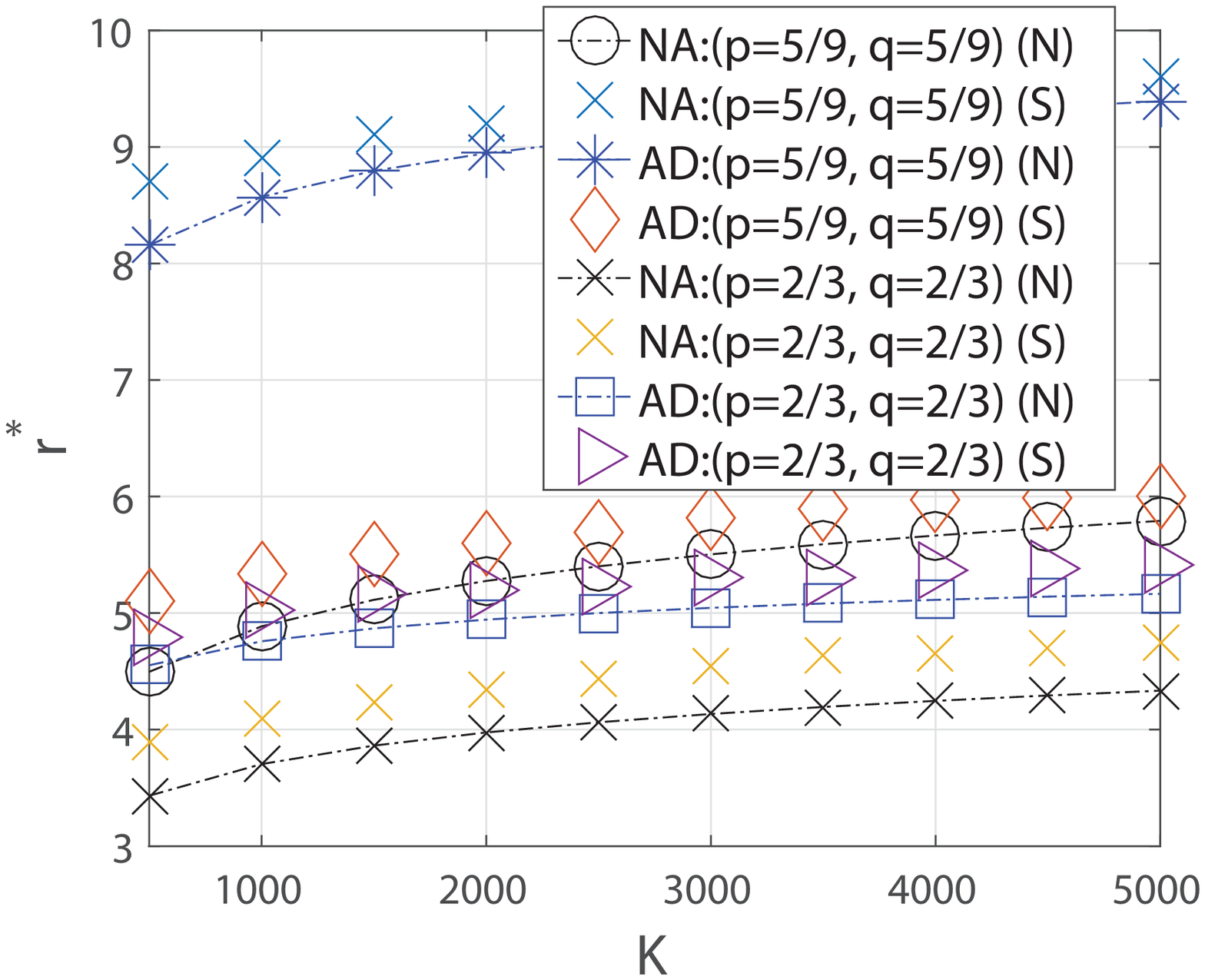}\label{fig:opt}}
\hspace{-0.19cm}
\subfigure[Detection with varying $p$.]{\includegraphics[width=0.5\columnwidth]{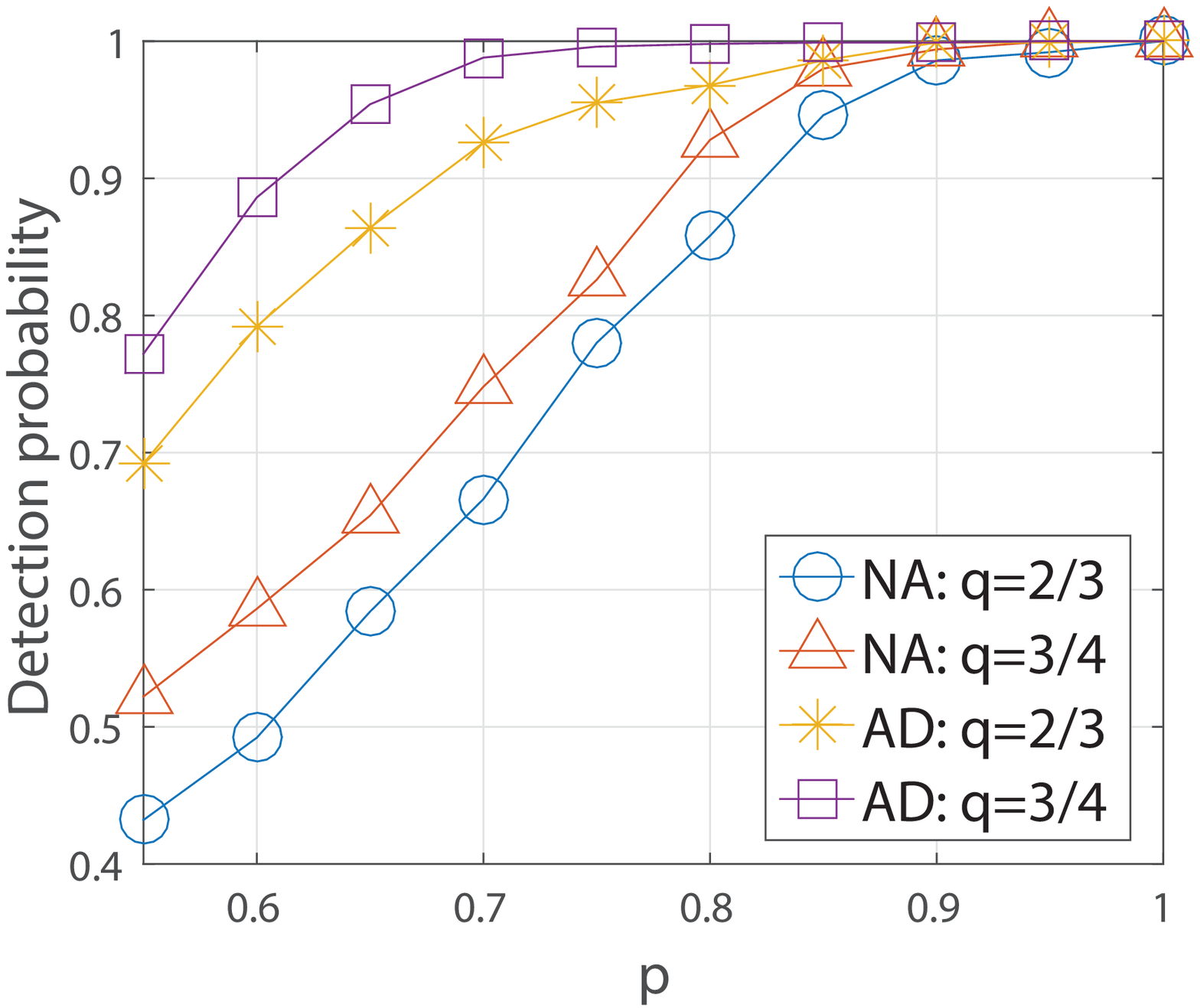}\label{fig:NAR}}
\hspace{-0.19cm}
\subfigure[Detection with varying $q$.]{\includegraphics[width=0.5\columnwidth]{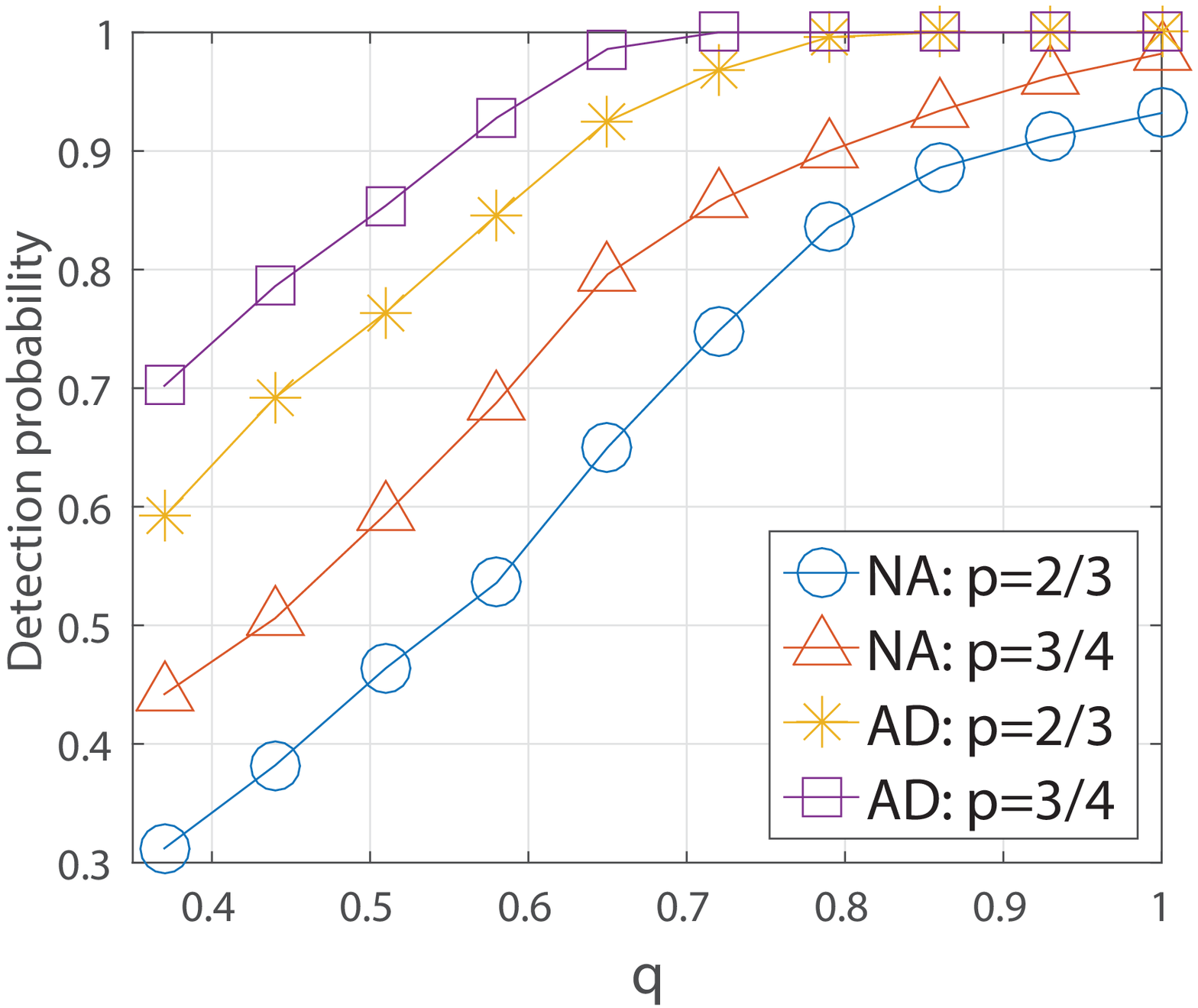}\label{fig:regularq}}
\hspace{-0.19cm}
\subfigure[Detection with varying $K$.]{\includegraphics[width=0.5\columnwidth]{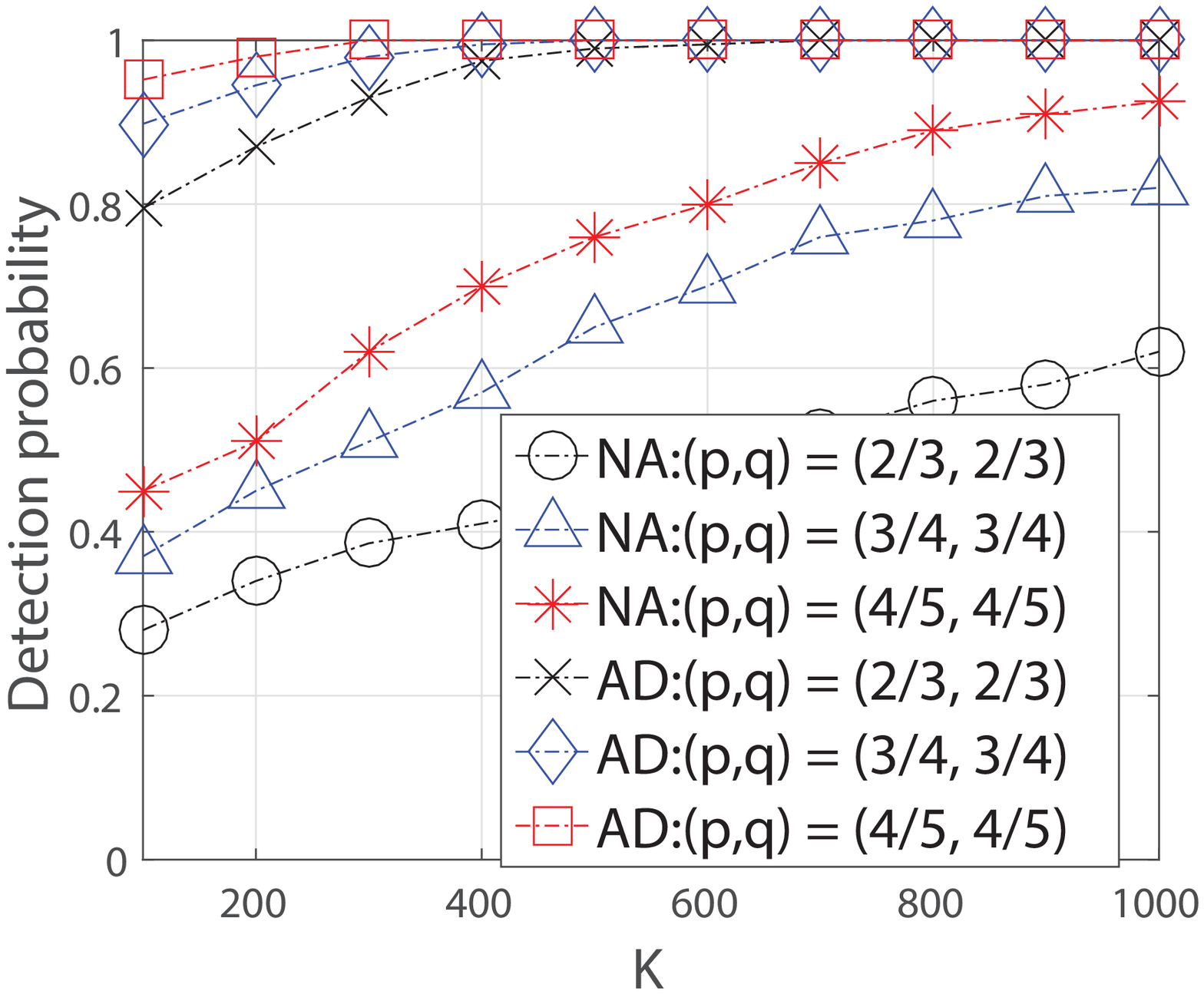}\label{fig:RGAD}}
\hspace{-0.19cm}
\caption{Results of regular trees (NA: Non-adaptive and AD: Adaptive): (a) Best $r^\star$ with varying budget $K$ (N:Numerical, S:Simulation), (b) Detection probabilities with varying $p$ under fixed $q$, (c) Detection probabilities with varying $q$ under fixed $p$, and (d) Detection probabilities versus budget $K$.}
\label{fig:regular}
\end{center}
\end{figure*}

\begin{figure*}[t!]
\begin{center}
\subfigure[Irregular Tree.]{\includegraphics[width=0.5\columnwidth]{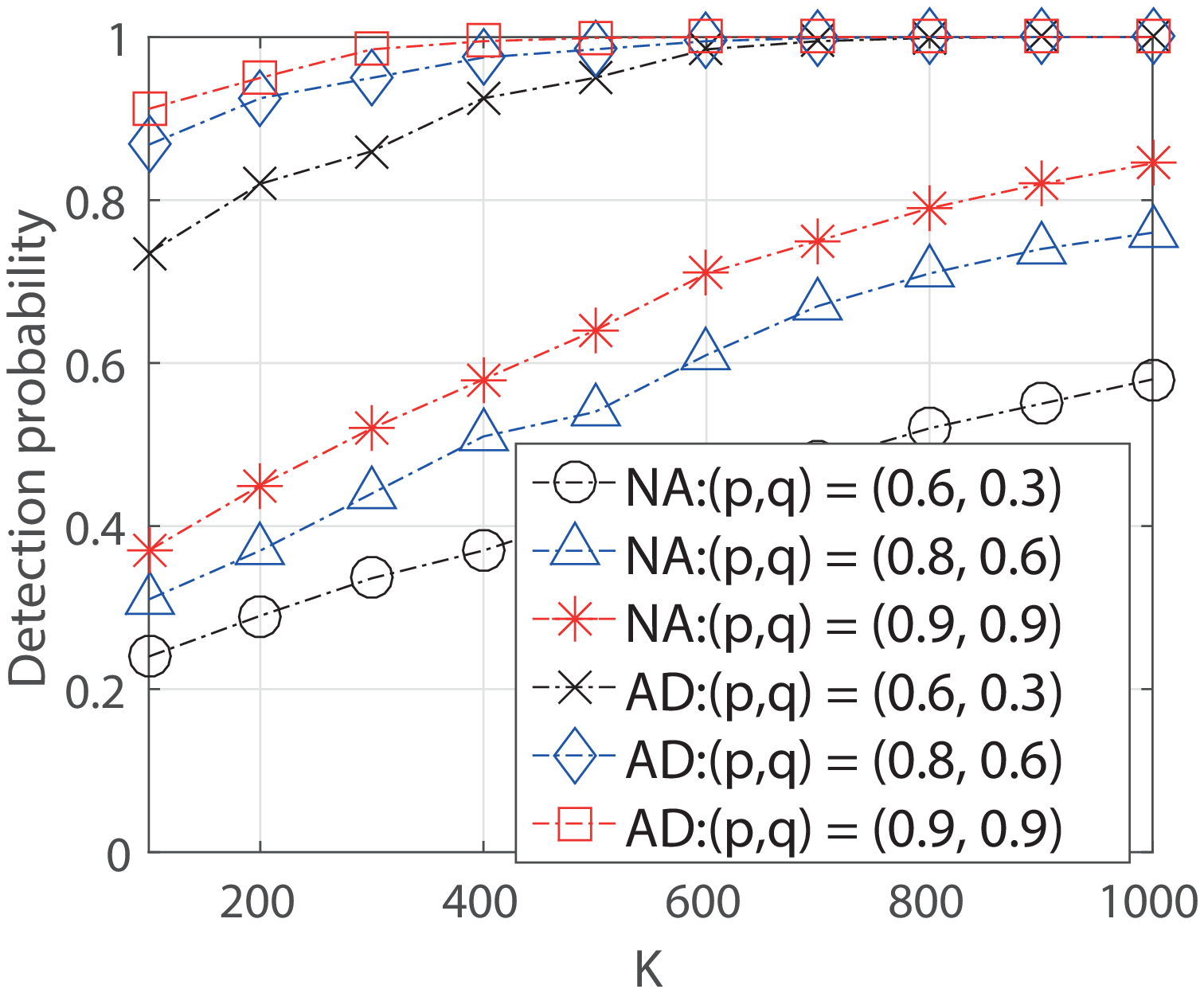}\label{fig:irre}}
\hspace{-0.19cm}
\subfigure[ER.]{\includegraphics[width=0.5\columnwidth]{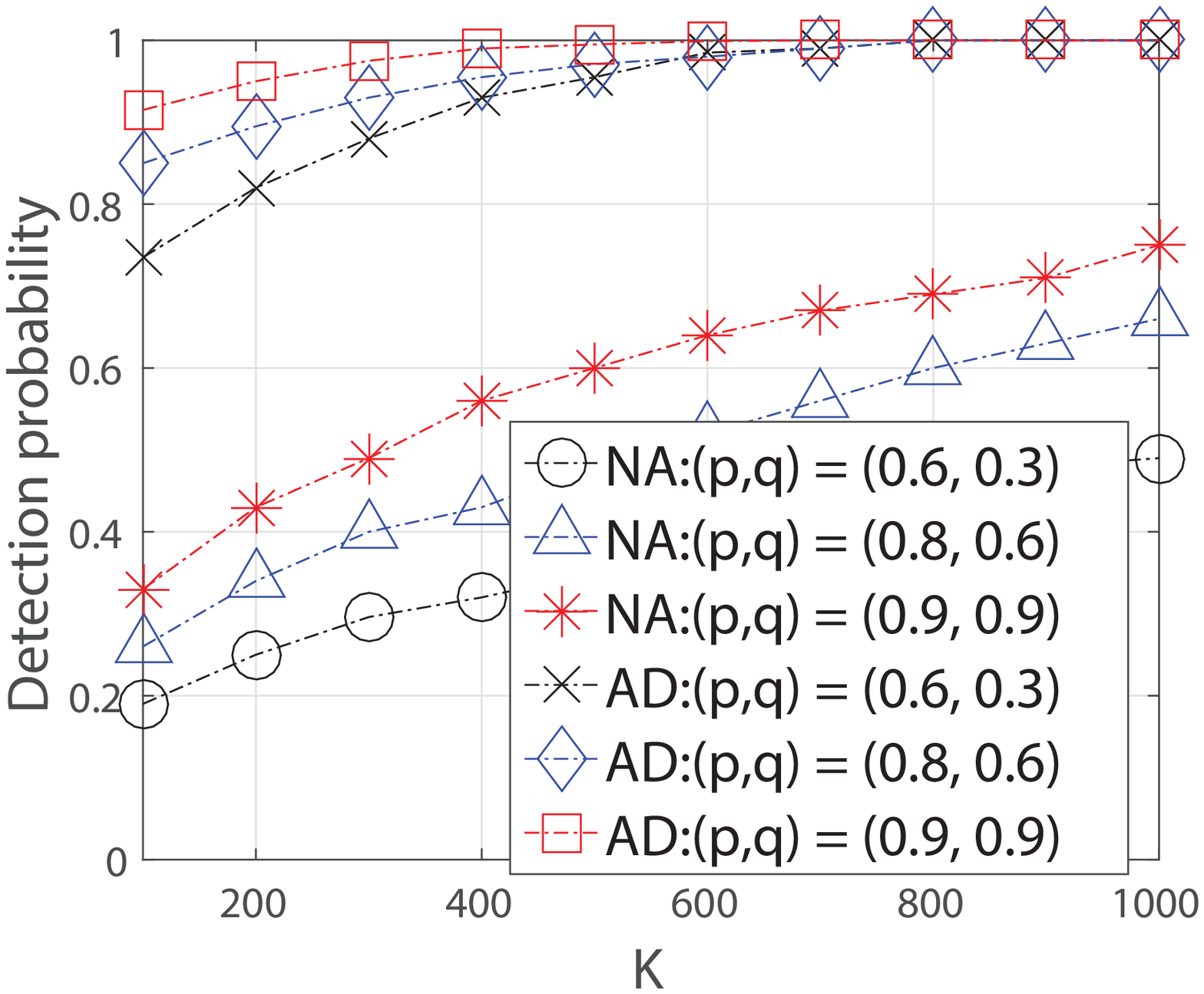}\label{fig:ERAD}}
\hspace{-0.19cm}
\subfigure[Scale Free.]{\includegraphics[width=0.5\columnwidth]{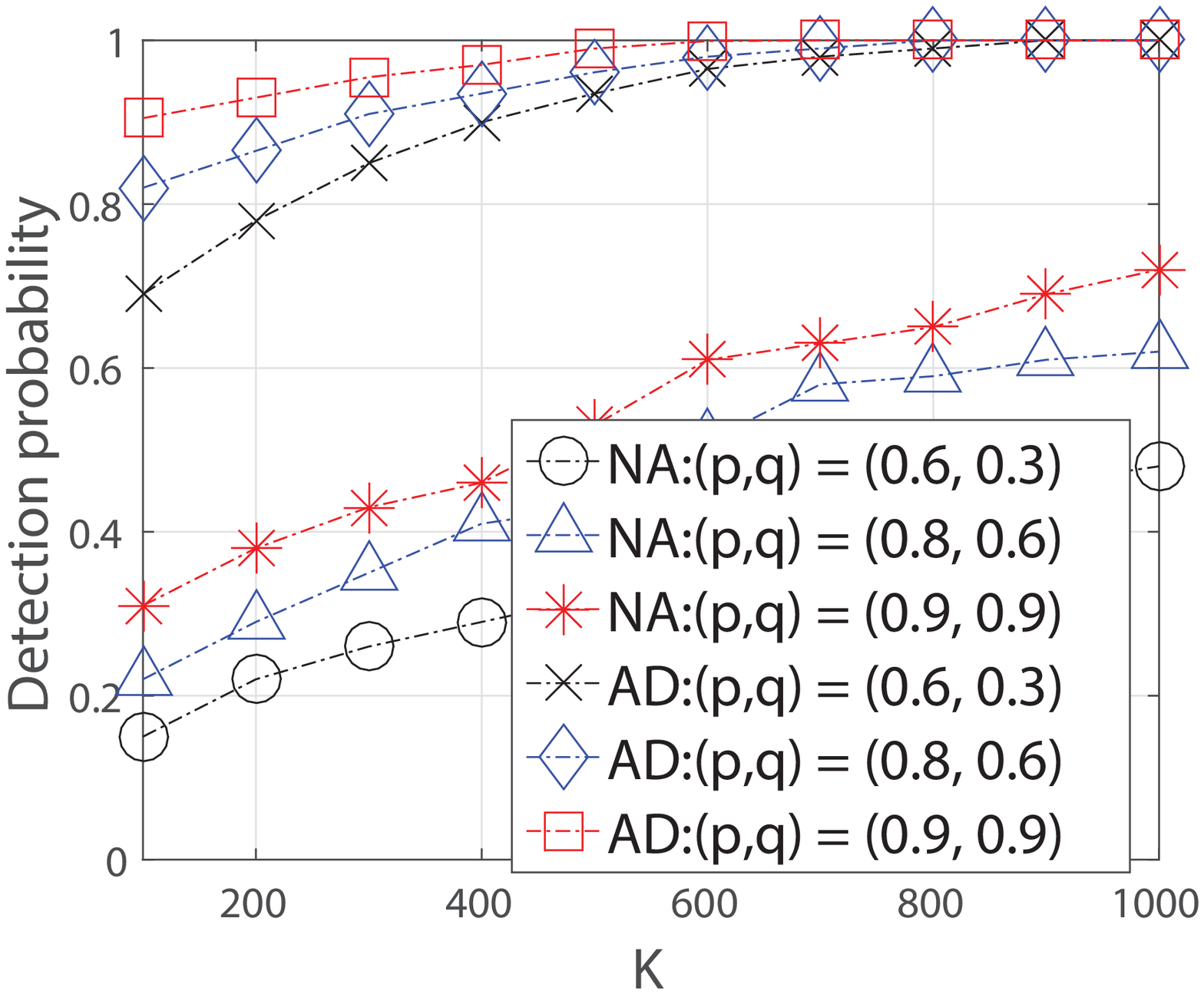}\label{fig:SFAD}}
\caption{Result of random graphs: Detection probabilities of (a) Irregular tree, (b) ER random graph, and (c) SF graph with varying budget $K$.}
\label{fig:synthetic}
\end{center}
\end{figure*}

\begin{figure*}[t!]
\begin{center}
\subfigure[Facebook graph.]{\includegraphics[width=0.5\columnwidth]{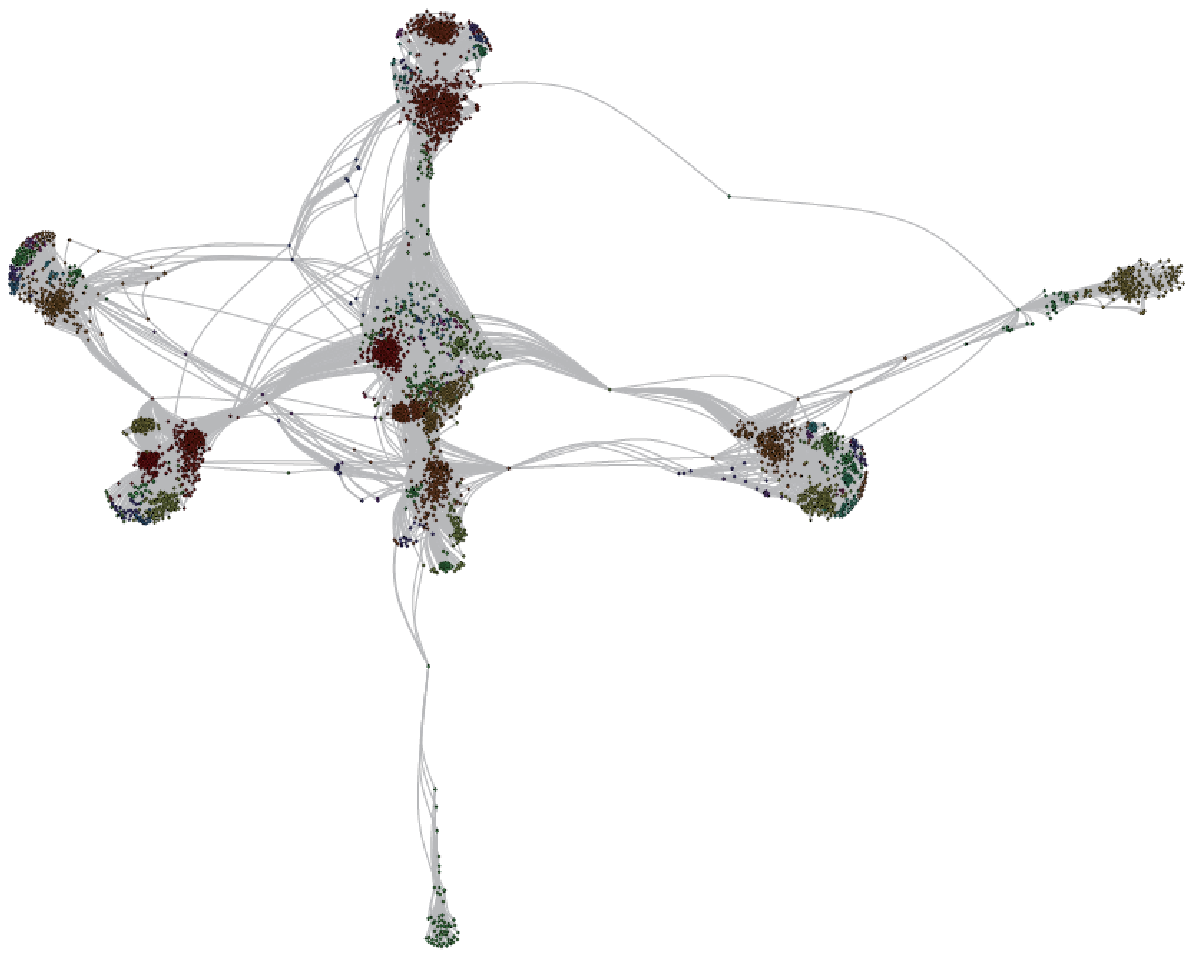}\label{fig:facebook}}
\subfigure[Facebook.]{\includegraphics[width=0.5\columnwidth]{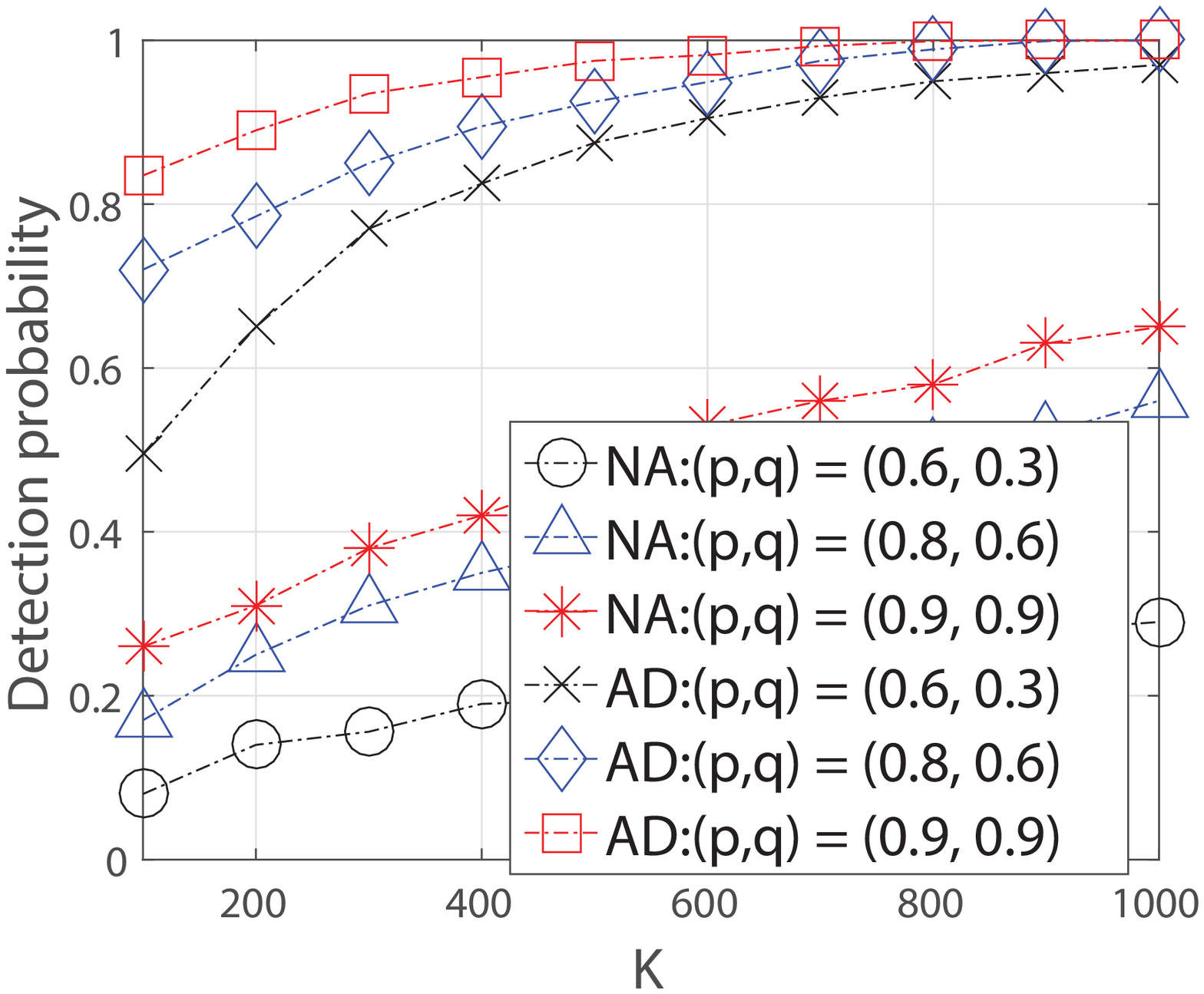}\label{fig:FBAD}}
\subfigure[Wiki-vote.]{\includegraphics[width=0.5\columnwidth]{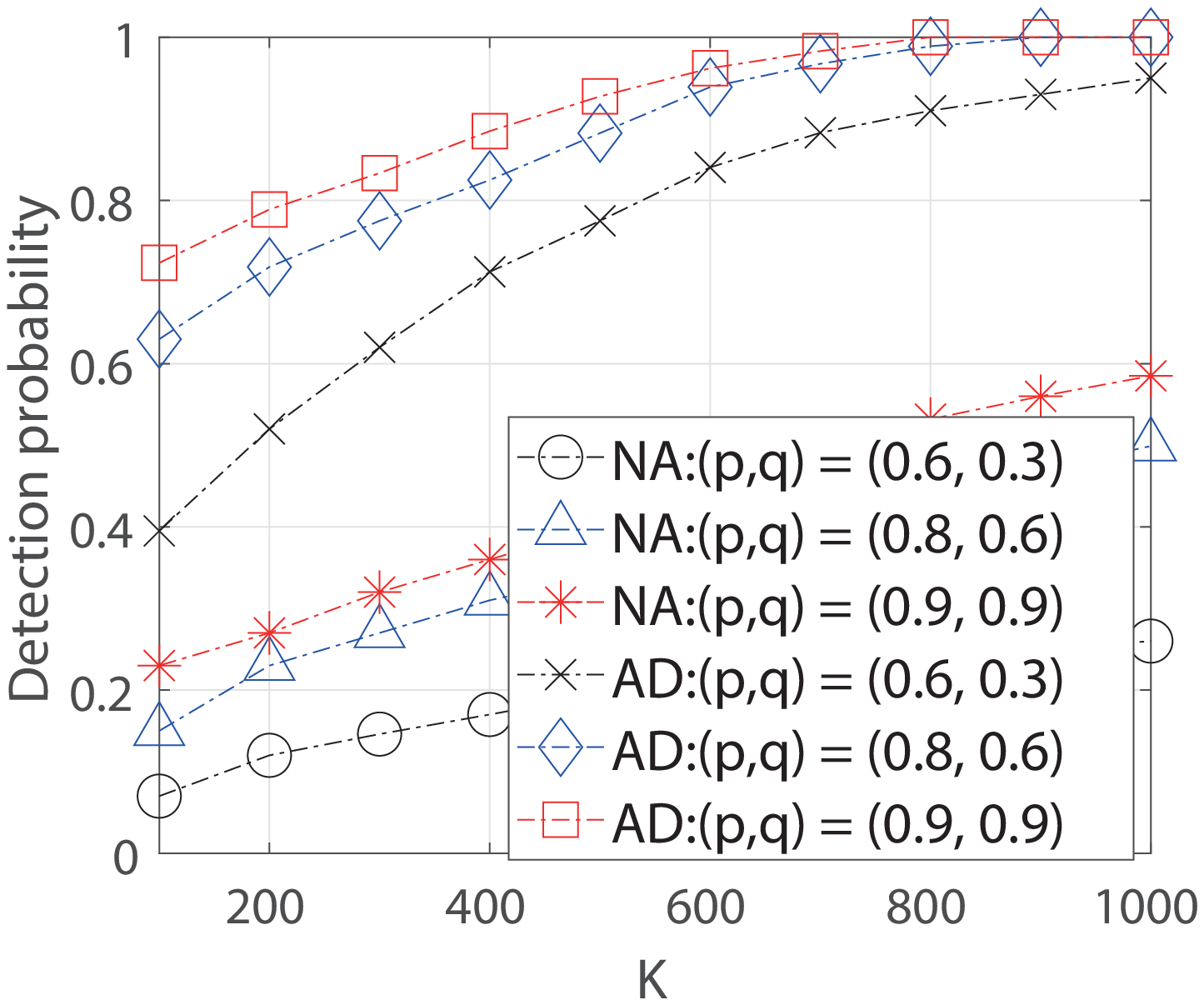}\label{fig:WIKI}}
\caption{Result of real world graphs: (a) Facebook graph, (b) Detection probabilities of Facebook graph with varying budget $K$, and (c) Wiki-vote graph with varying budget $K$.}
\label{fig:general}
\end{center}
\vspace{-0.39cm}
\end{figure*}

\section{Simulation Results}
\label{sec:numerical}

In this section, we will provide simulation results of our two proposed
algorithms over three types of graph topologies: $(i)$ regular trees, $(ii)$
random graphs, and $(iii)$ real world graphs, respectively. We propagate an information from
a randomly chosen source up to 400 infected nodes, and plot the
detection probability from 200 iterations.


\smallskip
\noindent{\bf \em (i) Regular trees.} For the regular tree, 
we first obtain the best $r^\star$ in each
theorems of both querying schemes as in Fig.\ref{fig:regular} (a). In this result, we
plot the obtained $r^\star$ without a flooring and run the simulation 100 iterations to obtain the mean under
the $d=3$ for both querying schemes\footnote{
In order to verify the query effect, we consider the simplest setting in the regular tree.}. We see that if we use the parameters $p=q=2/3$ then
the best $r^\star$ is in $[4, 6]$ but if the parameters are decreased by $p=q=4/9$ then
the algorithms use higher value of $r^\star$. Especially, we check that $r^\star = 10$
for the adaptive querying scheme due to increasing of the untruthfulness for direction answers.
Next, we obtain the detection probability for both querying schemes with $d=3$ that is
the ratio of the number of correct detections and iterations as varying two important
parameters $p$ and $q$, respectively. 
We first obtain the detection probabilities when one parameter is fixed ($q$ for the {\bf NA}-querying and $p$ for the
{\bf AD}-querying) under the given budget $K=200$ for the {\bf NA}-querying and $K=100$ for the {\bf AD}-querying
as in Fig.\ref{fig:regular} (b) and Fig.\ref{fig:regular} (c), respectively.
We check that if one parameter goes to one, then the detection probability also goes to one
regardless for the other parameter. This is because the truthfulness for the identity question is enough
to find the source in the {\bf NA}-querying if it is sufficiently large and the truthful direction query
can enlarge the detection for the {\bf AD}-querying, respectively. 
Finally, we see that {\bf NA}-querying needs more number of
budget to achieve the same target detection probability under the same parameter $(p,q)$ in
Fig.\ref{fig:regular} (d). 






\smallskip
\noindent{\bf \em (ii) Random graphs.}
As random graphs, we consider irregular tree (random tree), \emph{Erd\"{o}s-R\'{e}nyi} (ER) and Scale-Free (SF) graphs. For the irregular tree, we consider that the degree is upper bounded by $d_{max}=10.$ To generate the irregular tree, we use a Galton-Watson branch process with $d_{max}$. After the diffusion of information over the tree, we compute the RC for each infected node and select the maximal one to perform both query processes.
In the ER graph, we choose its parameter so that the average degree by 4
for 2000 nodes. In the SF graph, we choose the parameter so that the
average ratio of edges to nodes by 1.5 for 2000 nodes. It is known that obtaining MLE
is hard for the graphs with cycles, which is $\sharp$P-complete. Due to
this reason, we first construct a diffusion tree from the Breadth-First
Search (BFS) as used in \cite{shah2010}: Let $\sigma_{v}$ be the
infection sequence of the BFS ordering of the nodes in the given graph,
then we estimate the source $v_{\tt bfs}$ that solves the following:
\begin{equation}\label{eqn:BFS}
\begin{split}
  v_{\tt bfs}=\arg \max_{v\in
    G_{N}}\mathbb{P}(\sigma_{v}|v)R(v,T_{b}(v)),
\end{split}
\end{equation}
where $T_{b}(v)$ is a BFS tree rooted at $v$ and the rumor spreads along
it and $\mathbb{P}(\sigma_{v}|v)$ is the probability that generates the
infection sequence $\sigma_{v}$. Then, by using those selected nodes, we perform our algorithms
by changing the two parameters $(p,q)$. For the {\bf NA}-querying, due to the loop
in the general graph, the wrong answer for the direction question can be
a consistent edge. To avoid this issue, we use BFS tree to count the number of
these edges for all the nodes in the candidate set.
Fig.~\ref{fig:synthetic} (a), Fig.~\ref{fig:synthetic} (b) and Fig.~\ref{fig:synthetic} (c)
show the detection probabilities with varying $K$ for {\bf NA}-querying and
{\bf AD}-querying, where we observe similar trends to those in
regular trees. We check that how much direction information can enlarge the
detection performance compared to the case in \cite{Choi17} where does not
use the direction information in {\bf NA}-querying scheme. We also check that
the adaptivity gives the chance to find the source more efficiently in the
sense of using the queries. Further, we see that the detection performance for ER random graph is higher than that of SF graph. This indicates that the symmetric network topology (nodes have similar connectivity) with small diameter is good for finding the source by our hop-distance based model because the probability that the true source is in the candidate set will be large.

\smallskip
\noindent{\bf \em (iii) Real world graphs.}
Finally, we consider two real world graphs such as Facebook graph and WiKi-vote network in the simulation result. We use the
Facebook ego network as depicted in Fig.~\ref{fig:general} (a) in \cite{NIPS2012}, which is an undirected graph
consisting of 4039 nodes and 88234 edges, where each edge corresponds to
a social relationship (called FriendList) and the diameter is 8 hops. For the WiKi-vote network, we use the data in \cite{Wiki}, which generates 7115 nodes and 103689 edges, and the diameter is 7 hops. We
perform the same algorithm used for random graphs based on the BFS
heuristic and show the results in Fig.~\ref{fig:general} (b) and Fig.~\ref{fig:general} (c).
We see that the {\bf AD}-querying is quite powerful in the inferring
the source for both graphs. This is because the network as in Fig.~\ref{fig:general} (a) consists of small number of clusters containing many users. This leads to the result that the (shortest) distance of nodes in the graph is not large and interactive using the direction querying answer gives high chance to track the source in the network even though there are huge number of nodes. Further, we see that the {\bf AD}-querying is still powerful for the wiki-vote network as in Fig.~\ref{fig:general} (c).




\section{Conclusion}
\label{sec:conclusion}
In this paper, we considered querying for the information source inference problem in both non-adaptive and adaptive setting. We have provided some
theoretical performance guarantees when the underlying network has
regular tree structure. We obtained the answer for the fundamental
question of how much benefit adaptiveness in querying provides in finding the
source with analytical characterization in presence of individuals' untruthfulness by two proposed querying algorithms and information theoretical techniques to achieve the target probability
when the truth probabilities are homogeneous in the respondents. 
We also performed various simulation based on these algorithms.






\bibliographystyle{IEEEtran}
 \bibliography{reference}

\begin{thebibliography}{10}
\providecommand{\url}[1]{#1}
\csname url@samestyle\endcsname
\providecommand{\newblock}{\relax}
\providecommand{\bibinfo}[2]{#2}
\providecommand{\BIBentrySTDinterwordspacing}{\spaceskip=0pt\relax}
\providecommand{\BIBentryALTinterwordstretchfactor}{4}
\providecommand{\BIBentryALTinterwordspacing}{\spaceskip=\fontdimen2\font plus
\BIBentryALTinterwordstretchfactor\fontdimen3\font minus
  \fontdimen4\font\relax}
\providecommand{\BIBforeignlanguage}[2]{{%
\expandafter\ifx\csname l@#1\endcsname\relax
\typeout{** WARNING: IEEEtran.bst: No hyphenation pattern has been}%
\typeout{** loaded for the language `#1'. Using the pattern for}%
\typeout{** the default language instead.}%
\else
\language=\csname l@#1\endcsname
\fi
#2}}
\providecommand{\BIBdecl}{\relax}
\BIBdecl

\bibitem{Choi17}
J.~Choi, S.~Moon, J.~Woo, K.~Son, J.~Shin, and Y.~Yi, ``{Rumor Source Detection
  under Querying with Untruthful Answers},'' in \emph{{Proc. IEEE INFOCOM}},
  2017.

\bibitem{Choi18}
J.~Choi and Y.~Yi, ``{Necessary and Sufficient Budgets in Information Source
  Finding with Querying: Adaptivity Gap},'' in \emph{{Proc. IEEE ISIT}}, 2018.

\bibitem{Kai2016}
K.~Zhu and L.~Ying, ``{Information Source Detection in Network: Possiblity and
  Impossibility Results},'' in \emph{{Proc. IEEE INFOCOM}}, 2016.

\bibitem{shah2010}
D.~Shah and T.~Zaman, ``{Detecting Sources of Computer Viruses in Networks:
  Theory and Experiment},'' in \emph{{Proc. ACM SIGMETRICS}}, {2010}.

\bibitem{shah2012}
------, ``{Rumor Centrality: A Universal Source Estimator},'' in \emph{{Proc.
  ACM SIGMETRICS}}, 2012.

\bibitem{zhu2013}
K.~Zhu and L.~Ying, ``{Information Source Detection in the SIR Model: A Sample
  Path Based Approach},'' in \emph{Proc. IEEE Information Theory and
  Applications Workshop (ITA)}, 2013.

\bibitem{Luo2013}
W.~Luo and W.-P. Tay, ``{Finding an infection source under the SIS model},'' in
  \emph{{Proc. IEEE International Conference on Acoustics, Speech and Signal
  Processing (ICASSP)}}, 2013.

\bibitem{bubeck2014}
S.~Bubeck, L.~Devroye, and G.~Lugosi, ``{Finding Adam in random growing
  trees},'' in \emph{arXiv:1411.3317}, 2014.

\bibitem{Chang2015}
B.~Chang, F.~Zhu, E.~Chen, and Q.~Liu, ``{Information Source Detection via
  Maximum A Posteriori Estimation},'' in \emph{{Proc. IEEE ICDM}}, 2015.

\bibitem{farajtabar2015}
M.~Farajtabar, M.~Gomez{-}Rodriguez, N.~Du, M.~Zamani, H.~Zha, and L.~Song,
  ``{Back to the Past: Source Identification in Diffusion Networks from
  Partially Observed Cascades},'' in \emph{Proc. AISTATS}, 2015.

\bibitem{Zhang2014}
Z.~Wang, W.~Dong, W.~Zhang, and C.~W. Tan, ``{Rumor source detection with
  multiple observations: fundamental limits and algorithms},'' in \emph{{Proc.
  ACM SIGMETRICS}}, 2014.

\bibitem{dong2013}
W.~Dong, W.~Zhang, and C.~W. Tan, ``{Rooting Out the Rumor Culprit from
  Suspects},'' in \emph{Proc. IEEE ISIT}, 2013.

\bibitem{Fujii19}
K.~Fujii and S.~Sakaue, ``{Beyond Adaptive Submodularity: Approximation
  Guarantees of Greedy Policy with Adaptive Submodularity Ratio},'' in
  \emph{ICML}, 2019.

\bibitem{Singer16}
Y.~Singer, ``{Influence maximization through adaptive seeding},'' \emph{{ACM
  SIGecom Exchanges}}, vol.~15, no.~1, pp. 32--59, 2016.

\bibitem{Sewoong16}
A.~Khetan and S.~Oh, ``{Achieving Budget-optimality with Adaptive Schemes in
  Crowdsourcing},'' in \emph{NIPS}, 2016.

\bibitem{Yun14}
S.~Yun and A.~Proutiere, ``{Community Detection via Random and Adaptive
  Sampling},'' \emph{{JMLR}}, vol.~35, pp. 1--38, 2014.

\bibitem{shah2010tit}
D.~Shah and T.~Zaman, ``{Rumors in a Network: Who's the Culprit?}'' \emph{{IEEE
  Transactions on Information Theory}}, vol.~57, no.~8, pp. 5163--5181, August
  2011.

\bibitem{zhu2014}
K.~Zhu and L.~Ying, ``{A robust information source estimator with sparse
  observations},'' in \emph{Proc. IEEE INFOCOM}, 2014.

\bibitem{Choi2016}
J.~Choi, S.~Moon, J.~Shin, and Y.~Yi, ``{Estimating the Rumor Source with
  Anti-Rumor in Social Networks},'' in \emph{{Proc. IEEE ICNP Workshop on
  Machine Learning}}, 2016.

\bibitem{Leng2014}
W.~Luo, W.~P. Tay, and M.~Leng, ``{How to Identify an Infection Source With
  Limited Observations},'' \emph{{IEEE Journal of Selected Topics in Signal
  Processing}}, vol.~8, no.~4, pp. 586--597, August 2014.

\bibitem{Kumar2017}
A.~Kumar, V.~S. Borkar, and N.~Karamchandani, ``{Temporally Agnostic
  Rumor-Source Detection},'' \emph{{IEEE Transactions on Signal and Information
  Processing over Networks}}, vol.~3, no.~2, pp. 316--329, February 2017.

\bibitem{Jiao2018}
J.~Jiang, S.~Wen, S.~Yu, B.~Liu, Y.~Xiang, and W.~Zhou, ``{Identifying
  Propagation Source in Time-Varying Networks},'' \emph{{Malicious Attack
  Propagation and Source Identification, Springer}}, pp. 2850--2865, November
  2018.

\bibitem{ICDM17}
B.~A. Prakash, J.~Vreeken, and C.~Faloutsos, ``{Efficiently Spotting the
  Starting Points of an Epidemic in a Large Graph},'' in \emph{{Proc. ICDM}},
  2017.

\bibitem{ZhuAAAI17}
K.~Zhu, Z.~Chen, and L.~Ying, ``{Catch'Em All: Locating Multiple Diffusion
  Sources in Networks with Partial Observations},'' in \emph{{Proc. Association
  for the Advancement of Artificial Intelligence (AAAI)}}, 2017.

\bibitem{JI17}
F.~Ji and W.~P. Tay, ``{An Algorithmic Framework for Estimating Rumor Sources
  With Different Start Times},'' \emph{{IEEE Transaction on Signal
  Processing}}, vol.~65, pp. 2517--2530, 2017.

\bibitem{Jaing15}
J.~Jaing, S.~Wen, S.~Yu, Y.~Xiang, and W.~Zhou, ``{An Approach on the
  Multi-Source Identification of Information Diffusion},'' \emph{{IEEE
  Transactions on Information Forensics and Security}}, vol.~10, pp.
  2616--2626, 2015.

\bibitem{Khim14}
J.~Khim and P.-L. Loh, ``{Confidence Sets for Source of a Diffusion in Regular
  Trees},'' in \emph{arXiv:1510.05461}, 2015.

\bibitem{Sujay2012}
P.~Netrapalli and S.~Sangavi, ``{Learning the Graph of Epidemic Cascades},'' in
  \emph{{Proc. ACM SIGMETRICS}}, 2012.

\bibitem{Jae16}
\BIBentryALTinterwordspacing
[Tech] J. Choi, S. Moon, J. Woo, K. Son, J. Shin and Y. Yi, ``Information
  Source Finding in Networks: Querying with Budgets". [Online]. Available:
  \url{https://www.dropbox.com/s/87o1yaiu5t9bvyg/JY__Information_Source_Finding_Supply.pdf?dl=0}
\BIBentrySTDinterwordspacing

\bibitem{NIPS2012}
J.~McAuley and J.~Leskovec, ``{Learning to discover social circles in ego
  networks},'' in \emph{{Proc. NIPS}}, 2012.

\bibitem{Wiki}
J.~Leskovec, D.~Huttenlocher, and J.~Kleinberg, ``{Predicting Positive and
  Negative Links in Online Social Networks.}'' in \emph{{Proc. WWW}}, 2010.

\end{thebibliography}

\end{document}